\documentclass[12pt]{article}

\usepackage{lineno}

\usepackage{hyperref}
\usepackage{pdfpages}
\usepackage[utf8]{inputenc}
\usepackage{authblk}
\usepackage{graphicx}
\usepackage{amssymb}
\usepackage{amsmath,amsthm}
\usepackage{url}
\usepackage{color}
\usepackage[numbers,comma,compress]{natbib}
\usepackage[defaultfam,light,tabular,lining]{montserrat} 

\usepackage{setspace}
\usepackage{caption}
\usepackage{parskip}
\usepackage[utf8]{inputenc}

\usepackage[top=0.8in, bottom=1.6in, left=1in, right=1in]{geometry}
\title{Deciphering the global production network from cross-border firm transactions}

\date{}

\author[1*]{\small Neave O'Clery}
\author[2]{Ben Radcliffe-Brown}
\author[2]{Thomas Spencer}
\author[2]{Daniel Tarling-Hunter}

\affil[1]{Centre for Advanced Spatial Analysis, University College London}
\affil[2]{UK Foreign, Commonwealth and Development Office}
\affil[*]{Corr. Author: n.oclery@ucl.ac.uk}

\begin{document}

\maketitle

\begin{abstract}

Critical for policy-making and business operations, the study of global supply chains has been severely hampered by a lack of detailed data. Here we harness international firm-level transaction data covering 20m global firms, and 1 billion cross-border transactions, to infer key inputs for over 1200 products. Transforming this data to a directed network, we find that products are clustered into three large groups including textiles, chemicals and food, and machinery and metals. European industrial nations and China dominate critical intermediate products such as metals, common components and tools, while industrial complexity is highly correlated with embeddedness in densely connected supply chains. Both forward and backward linkages are predictive of country-product diversification patterns, with stronger overall evidence for backward (upstream) linkages. Finally, we find structural similarities with AIPNET, a reference network generated via LLM queries, and strong linkages between products identified in manually-mapped electric vehicle battery and semiconductor supply chains. 

\end{abstract}

\textbf{Teaser: Harnessing 1 billion cross-border inter-firm transactions to reconstruct the global supply chain production network.}

\textbf{Disclaimer: }Any views expressed are solely those of the author(s) and so cannot be taken to represent those of the Foreign Commonwealth and Development Office (FCDO), or to state FCDO Policy. This paper should therefore not be reported as representing the views of the FCDO, its Ministers or His Majesty's Government.

\clearpage

\onecolumn
\section*{Introduction}

As acutely highlighted by the global pandemic, limited insight into global supply chain dynamics can have serious and immediate consequences. However, until quite recently, supply chain analysis was limited to specific sectors or highly aggregate economy-wide transaction data which lacks detail and hence is not sufficiently accurate for many policy and research questions. Here we harness 'next generation' global inter-firm transaction data to infer detailed product inputs as the basis for constructing a product-level global supply chain network. 

Traditionally, flows and transactions between sectors in the economy have been captured via input-output (IO) tables \cite{leontief1986input}. These capture monetary flows between firms within defined industries, and are typically produced by national statistical agencies. Some countries, such as the US and Mexico, produce these at quite a detailed level (4 digit NAICS - or approx 400 industries). However, in most countries they are only available for very aggregate sectors (in the UK, for example, IO tables cover just 105 sectors). And, importantly, they don't capture cross border flows limiting their potential for global supply chain analysis. There have been a handful of efforts to build integrated input-output networks that include inter-country flows, but again these are extremely aggregate in nature. For example, the World Input-Output Dataset (WIOD) \cite{timmer2012world} has just 56 sectors and 40 countries (2010-14) and the OECD World Input-Output Table (WIOT) has just 36 sectors and 69 countries (1995-2020). 

In recent years there has been a huge push to better collect (and make available for research) firm-level transaction data which has until now been mainly available for individual countries \cite{pichler2023building}. These national datasets share common characteristics \citep{bacilieri2023firm}, and have been deployed to show that, for example, systemic risk in production networks is concentrated in firms that provide essential intermediate goods \cite{diem2022quantifying, diem2024estimating, inoue2024disruption}. This data is typically, however, collected by single countries (e.g., VAT data) or for a specific purpose (i.e., shipping data), and hence has very limited coverage when it comes to studying global supply chains. 

At the other end of the spectrum are efforts to manually reconstruct detailed supply chains for specific sectors. These studies collect data from a variety of sources including internal systems like procurement, inventory management, and logistics, as well as external sources such as supplier data and transportation information, to build a detailed picture of the key inputs needed to produce often strategically important goods. For example, there have been careful efforts to map out antibiotic \cite{klimek2023anatomy}, semi-conductor \cite{haramboure2023vulnerabilities} and critical minerals in electric vehicle battery \cite{UNCTAD} and defence \cite{girardi2023strategic} supply chains. However, these mapping efforts tend to focus on material inputs such as critical minerals and chemicals rather than machinery, tools and other necessary inputs such as cleaning equipment. And, critically, their fragmented nature makes them not suited to many questions which require a wider structural lens on supply chains inter-connectivity and inter-dependence. 

Here we aim to infer detailed 'product recipes', or the set of unique inputs needed in the production of a specific product. In other words, we aim to capture the input materials and tools that are specific to a particular product (and hence not common across many products, e.g., packaging). Distinct from two recent efforts to build a similar product network using firm-level transaction data from specific countries (Italy \cite{fessina2025product} and Japan \cite{inoue2024disruption}), we aim to capture product linkages in global supply chains. To do this, we harness global cross-border inter-firm transaction data via a UK government programme, GSCIP. The data, procured under license by GSCIP, was compiled by a commercial entity from a variety of sources, from customs to firm-owned data, and matched at a firm level. It contains information solely on cross-border inter-firm purchases, and does not capture local supply chain linkages.

To detect the set of inputs required to make a specific product, we compare the purchasing patterns of firms which make this input (producer firms) to those who do not. Specifically, we compute the ratio of producer firms to all firms who purchase each input. This latter step effectively down-weights common or ubiquitous inputs. This approach echoes that of \cite{o2021productive} in which the authors inferred typical product diversification paths by extracting the set of common products produced previously by countries which went on later to make a specific product. One can think of our approach as 'piecemeal' reconstruction of global product supply chains in that we independently estimate the set of input linkages for each product. In this way, we capture both inputs or ingredients, and complex tools and packaging. 

We represent our product recipes as a 'production network' in which nodes are products, and directed weighted edges represent the relative importance of the input (and not monetary flows such as those found in input-output tables). This network does not exhibit traditional chains with 'root nodes' or 'tree-like' structures, but is a complex representation of the inherent circularity of modern supply chains in which very complex goods such as mining equipment is purchased by firms making raw inputs. While there is no ‘ground truth’ data available, we have undertaken significant efforts to compare the network structure to other related efforts. We found structural similarities with an alternative AI-generated product network \cite{fetzer2024ai}, and a product supply chain dataset created by NAFTA to track product origins \cite{conconi2018final}. In addition, as a further quasi-validation step, we find non-random links in our network between nodes identified as belonging to two manually constructed supply chains - electric vehicle batteries \cite{UNCTAD} and semiconductors \cite{haramboure2023vulnerabilities}. 

A major focus of analytical efforts with respect to supply chains has been to better understand complex global inter-dependence structures and associated vulnerabilities \cite{blackhurst2005empirically}. In particular the pandemic and Iran war have drawn attention to  chokepoints in global supply chains, and the intense interdependencies that underlie modern production and supply systems \cite{farrell2019weaponized}. The structure of global supply chains thus encodes both vulnerabilities and the potential impact of shocks, ranging from localised shipping disruptions or natural disasters \cite{barrot2016input, inoue2019firm} to broad-based shocks such as the pandemic \cite{lafrogne2022supply} or financial crisis. Here, we infer information about chokepoints, interdependencies and dynamics in the global supply chain network by analysing the structure of our production network \cite{newman2011structure}. Deploying community detection, which captures the contours of dynamics such as shock propagation on production networks \cite{piccardi2018random}, we find that textiles’ supply chains are largely isolated from wider manufacturing, while chemicals are typically more inter-connected with food supply chains than machinery and metals. European industrial nations and China dominate nodes with high betweenness centrality, i.e., critical intermediate products in the network, while countries with high industrial complexity (proxied via the ECI metric of \cite{hidalgo2009building}) tend to export goods that have a high 'hub score' in the network, i.e., they are embedded in densely connected supply chains.

On a longer time-scale, information on linkages between products in a supply chain is also informative of potential product diversification paths of countries. We engage in a wide literature that studies the role of supply chains as a driver in the industrial diversification (or 'upgrading') of countries into higher value-added activities, particularly in resource-rich countries such as African nations with critical minerals \cite{hirschman1958strategy, gereffi1999commodity, kaplinsky2000globalisation}. Known as backward diversification, countries also move into 'upstream' goods such as machinery or components. Upstream goods include both basic inputs and highly complex and high-value goods such as chemicals, electronics and equipment produced by advanced manufacturing economies \cite{antras2012measuring, suganuma2016upstreamness}.

A wide array of mechanisms has been proposed for both upstream and downstream diversification in supply chains, with many focusing on the diffusion of knowhow via, for example, demands for quality upgrading, worker flows and increased competition \cite{blomstrom1998multinational, criscuolo2017relationship}. Hirschman (1958) \cite{hirschman1958strategy} suggested that the combination of a local market for the good and know-how would drive upstream diversification, particularly in developing countries. While these effects have been long documented in the case of supplier-customer relationships between local firms and co-located foreign multi-nationals \cite{javorcik2004does}, there are greater uncertainties about the potential for such linkages to induce upgrading when spanning large geographical and technological distances \cite{pietrobelli2011global}. Various empirical studies find evidence of upstream diversification \cite{bahar2019export, ndubuisi2021important} but have been limited to proxying international supply chain links from e.g., US input-output data. Deploying our network, which directly measures embeddedness of a country in global supply chains, we are able to predict country-product diversification patterns with high accuracy, with overall stronger results for diversification into upstream goods.

\section*{Results}

\textbf{We use global inter-firm transaction data covering over 20m firms and 1 billion transactions to construct the product network.} Firm level transaction data has been compiled from a variety of sources including customs and firm-held data, such as bill of lading (shipping) data. These sources have been integrated by a commercial provider using AI-based tools to, for example, disambiguate firm names and match them. The data covers transactions between firms located in different countries during the period 2021-23 and was made available to the research team via a UK government programme (GSCIP). GSCIP procured the firm transaction data from a commercial entity, and linked it to additional firm-level information from other sources, such as firm ownership structures (see SI for more detail). 

We have a number of firm and transaction attributes, a selection of which are shown in Figure 1 B, including dollar transactions for specific products between firm pairs. While our dataset does not include primary-source data for every country, as shown in Figure 1 C-D, the data has high coverage across developed economies including G7 countries, and those with high trade (exports plus imports) volumes such as China and Vietnam. It also has good coverage in some middle income countries such as India and Peru, but with lower numbers of firms overall in less economically active regions such as Africa. 
In SI Section 3.3, we show correlations of transaction inflows and outflows with official trade data on imports and exports (Comtrade), finding very high correlations at HS2 product level (0.97-0.99), good correlations for countries (0.62-0.65) with slightly better correlations for exports vs imports, and higher correlations for a subset of poorer countries relative to the most wealthy ones (0.67-69 vs 0.86-0.89). 

\begin{figure}[t!]
    \centering
  \includegraphics[width=1\textwidth]{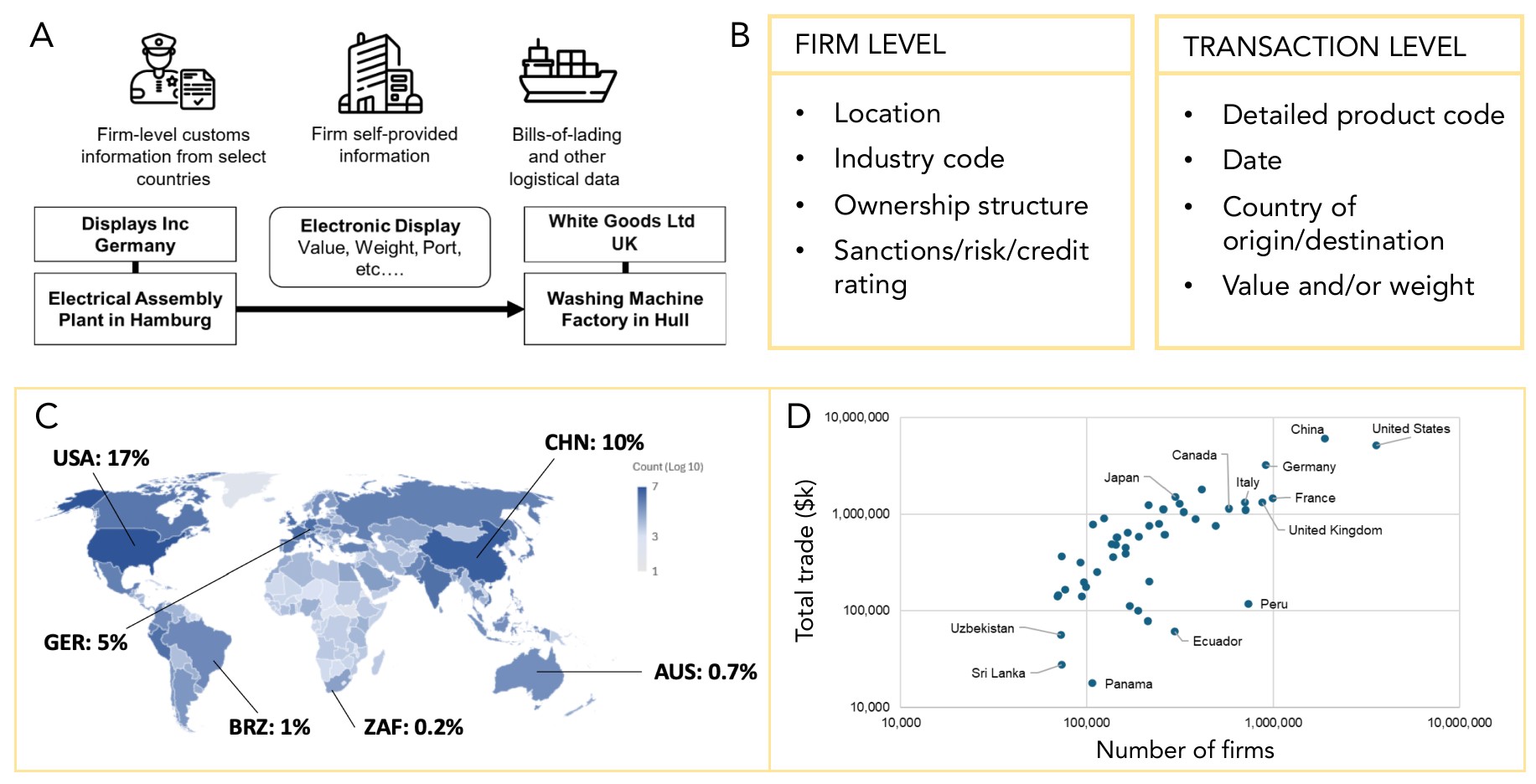}
  \caption{\textbf{We use global inter-firm transaction data covering over 20m firms and 1 billion transactions.} [A] Firm level transaction data has been compiled from a variety of sources. [B] This includes firm level attributes for 20m firms, combined with 1 billion inter-firm transactions. [C-D] The data has very good coverage of firms in developed economies, with lower coverage in less economically active regions such as Africa. Top 50 countries by firm count shown.}
  \label{fig1}
\end{figure}

\textbf{We infer the importance of each product as an input to any other product from the firm transaction data.} We aim to derive a metric to capture the 'importance' of product $i$ for the production of product $j$. In essence, we compute the 'excess' purchase of product $i$ by firms that produce product $j$, relative to the purchase of $i$ by 'an average firm'. In our toy example in Figure 2 B, we show that while aluminium panels are purchased very frequently by washing machine firms (relative to all firms), cardboard boxes are often purchased by all types of firms including washing machine firms and hence they are a very common input (and are thus down-weighted in our scheme).

More precisely, we set stringent conditions to identify firms that both produce and purchase specific products:
\begin{itemize}
\item A firm $k$ produces product $j$ if they sell a larger monetary value of $j$ than the average firm that sells product $j$.
\item A firm $k$ buys product $i$ if they purchase a larger monetary value of $i$ than the average firm that buys product $i$.
\end{itemize}
The set of all firms is denoted $S$. For each product $j$, we form a subset of firms that produce product $j$, denoted $S_j$. A subset of these buy product $i$, $S^i_j$. Finally, $S^i$ is the set of firms that purchase product $i$ (irrespective of what they produce). 

Our ratio is the share of firms who produce $j$ that also buy $i$ divided by the overall share of firms that buy $i$:
\begin{equation*}
A_{i,j}=\frac{|S^i_j|/|S_j|}{|S^i|/|S|}
\end{equation*}
If $A_{i,j}>1$, then product $i$ is more prevalent in the purchasing basket of firms that produce $j$ than the wider set of all firms. We can also set a minimum for the size of the set $|S^i_j|$ (the number of firms that both produce $j$ and buy $i$), which we refer to as parameter 'firmcount'. For low values of firmcount, we have few 'pieces of evidence' for the edge weight - and so for some applications we might want to set a higher firmcount threshold. This would make our network sparser, but we would have higher confidence in the edge weights as they are based on a higher number of signals from the data. In the analysis below, we typically show results for a range of thresholds on both the edge weight ($A_{i,j}$) and firmcount ($|S^i_j|$). 

As shown in Figure 2 B, we pick up both inputs or ingredients and complex tools. For example, for tomatoes, we capture everything from wooden stakes to fertilizer, harvesting machines and packaging.

\begin{figure}[t!]
    \centering
  \includegraphics[width=1\textwidth]{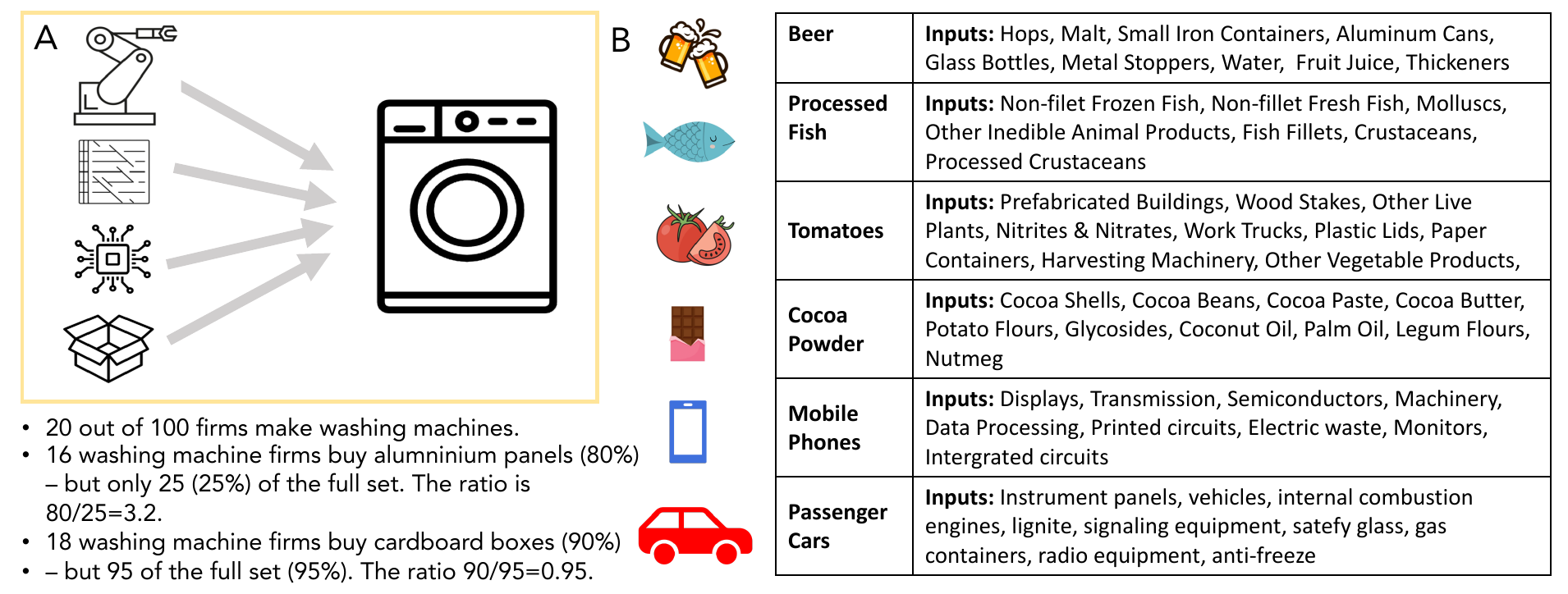}
  \caption{\textbf{We infer the importance of each product as an input to any other product from the firm transaction data.} [A] We compute the excess purchase of product $i$ by firms that produce product $j$, relative to the purchase of $i$ by 'an average firm'. In our toy example here we show that while aluminium panels are purchased very frequently by washing machine firms relative to all firms, cardboard boxes are purchases by all types of firms including washing machine firms and hence they are a very ubiquitous input (and are thus down-weighted in our scheme). [B] We pick up both inputs or ingredients and complex tools. For example, for beer, we capture everything from hops and malt to containers and bottle stoppers.}
  \label{fig2}
\end{figure}

\textbf{We transform the product input relations into a network, and find three main clusters representing machinery and metals, chemicals and food, and textiles.}
Our first task is to understand the complex structure of this set of product inputs. We focus initially on understanding the meso structure of the network, or to what extent do we have clusters of products which are closely interconnected in global supply chains, but disconnected from the wider set? It is at this intermediate scale that we get an initial sense of the overall structure of the data, and how those higher level contours might influence dynamics on the networks such as shocks or cascades \cite{lambiotte2015random}. 

To investigate this, we transform the product input relations into a directed weighted network, with edge weight stored in adjacency matrix $A$ (with entries $A_{i,j}$ as above) of size $n$ x $n$ for $n=1228$ products. In Figure 3 A, we observe dense edges within sectors (HS sections), shown in coloured blocks on the diagonal. As we increase the firmcount parameter, the edges become more sparse, but the edge weight threshold is less sensitive. In most of the analysis to follow we do not fix these parameters but show the results for a range of both parameters. In the few cases where this is impractical or unnecessary, we use largest weakly connected component of the network with edge threshold value 2 and firmcount value 2 (size $n=1207$).  

This network can be visualised using a 'spring' based visualisation algorithm (Force Atlas in Gephi was used here), as shown in Figure 3 B. We observe loose clustering of nodes according to HS 'Section', a high level classification of products into sectors, but with strong connections between sectors. We use a multi-scale community detection algorithm \cite{delvenne2010stability} to uncover the 'meso' structure or high-level organisation of the network into node 'communities'. These are clusters of products with strong internal connectivity - but relatively weaker connectivity to the rest of the network - and can be seen as an alternative classification of products based on supply chain connectivity. These structural patterns are highly informative, capturing for example the behaviour of cascades and shock propagation on the network \cite{fortunato2010community,lambiotte2015random}. Specifically, a shock which starts in a cluster is more likely to spread within the cluster first before transitioning or propagating to other parts of the network. 

\begin{figure}[t!]
    \centering
  \includegraphics[width=1\textwidth]{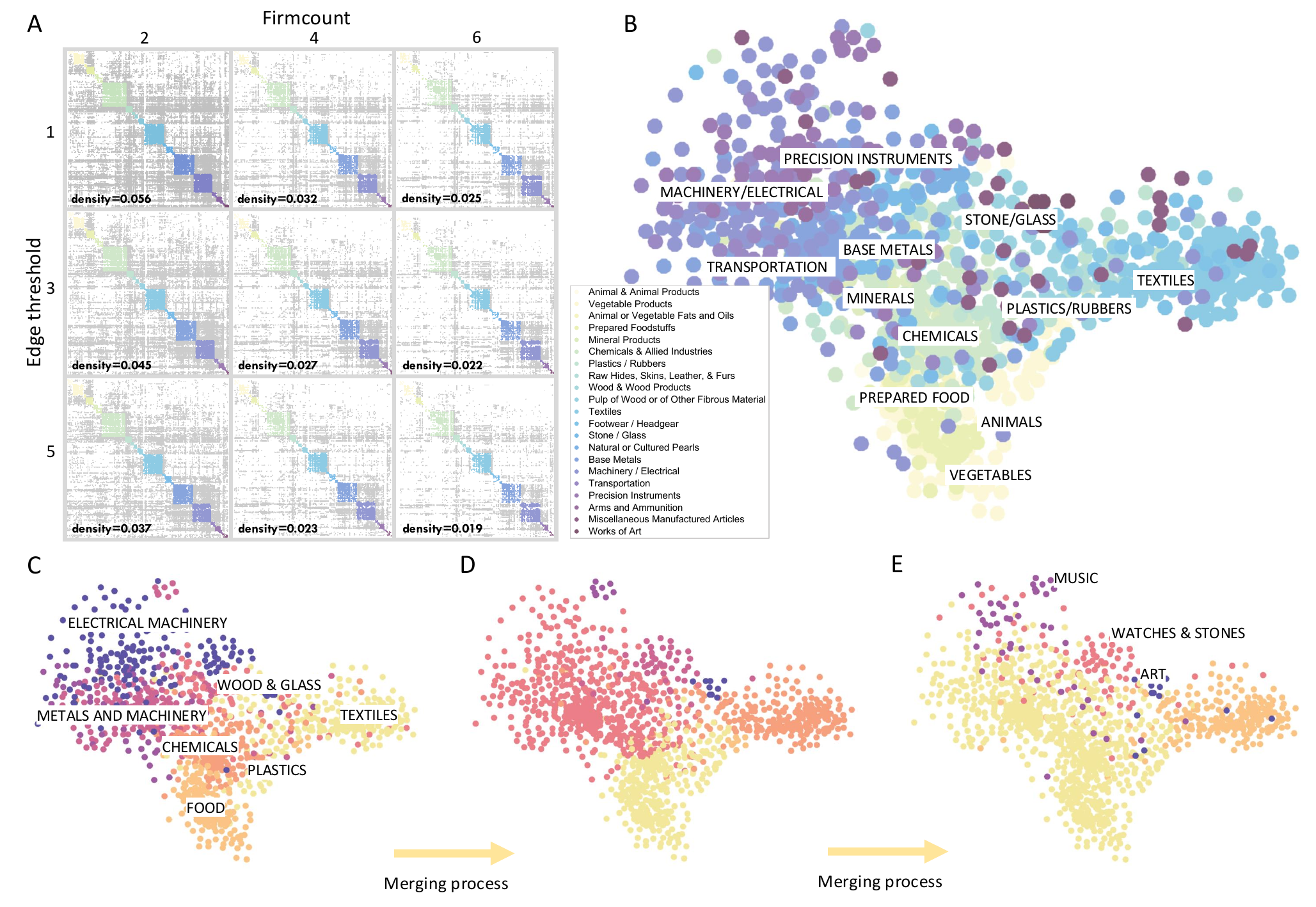}
  \caption{\textbf{We transform the product input relations into a network.} [A] The product supply chain network can represented by an adjacency matrix in which a value in entry $A_{i,j}$ denotes a non-zero edge weight between node (product) $i$ and node $j$. We observe dense edges within sectors (HS sections), shown in coloured blocks on the diagonal. As we increase the firm-count parameter, the edges become more sparse, but the edge weight threshold has no major effect. [B] The network can also be visualised using a visualisation algorithm (Force Atlas was used here). [C-E] We use a multi-scale community detection algorithm \cite{delvenne2010stability}  to uncover the 'meso' structure of the network. As the algorithm effectively 'forces' communities to merge into larger clusters, we find that textiles is highly isolated in the sense that it remains unmerged over many scales. Overall, we find that the product network is broadly segregated into three main clusters representing machinery and metals, chemicals and food, and textiles.}
  \label{fig3}
\end{figure}

While there are many distinct approaches to community detection suited to different applications \cite{fortunato2010community}, in our case, we employ a multi-scale algorithm based on random walkers on a network \cite{delvenne2010stability}. As walkers roam on the network, jumping from node to node based on probabilities derived from the directed edge weights, the algorithm detects larger and larger clusters. Hence, as 'time' increases, the algorithm typically detects fewer but larger communities. We can learn from this process: as the algorithm effectively 'forces' communities to merge into larger clusters (although this process is not strictly hierarchical), the 'time' at which they merge (or refuse to merge) is highly informative about their relative inter-connectivity. In other words, those that merge earlier are more connected, but those that remain unmerged for longer times are more isolated and disconnected from the wider network. 

Here, in Figure 3 C-E, we show three node partitions corresponding to three distinct 'times' or scales (see SI Figure S3 for more details and further analysis). The merging process suggests that textiles is highly isolated in the sense that it remains unmerged across all three partitions shown. We find that initially (Figure 3 C) machinery is split across two clusters - one focused on electrical goods and the other on metal-based tools (and metals themselves). As the algorithm detects larger clusters over longer times, food merges with plastics and then chemicals, and both machinery clusters merge. Then, these two groupings merge - but textiles and some other small clusters (musical products, watches, and art) remain unmerged and are thus highly isolated with weak connectivity to the wider network. Overall, we find that the product network is broadly segregated into three main clusters representing machinery and metals, chemicals and food, and textiles. 

\begin{figure}[t!]
  \centering
  \includegraphics[width=1\textwidth]{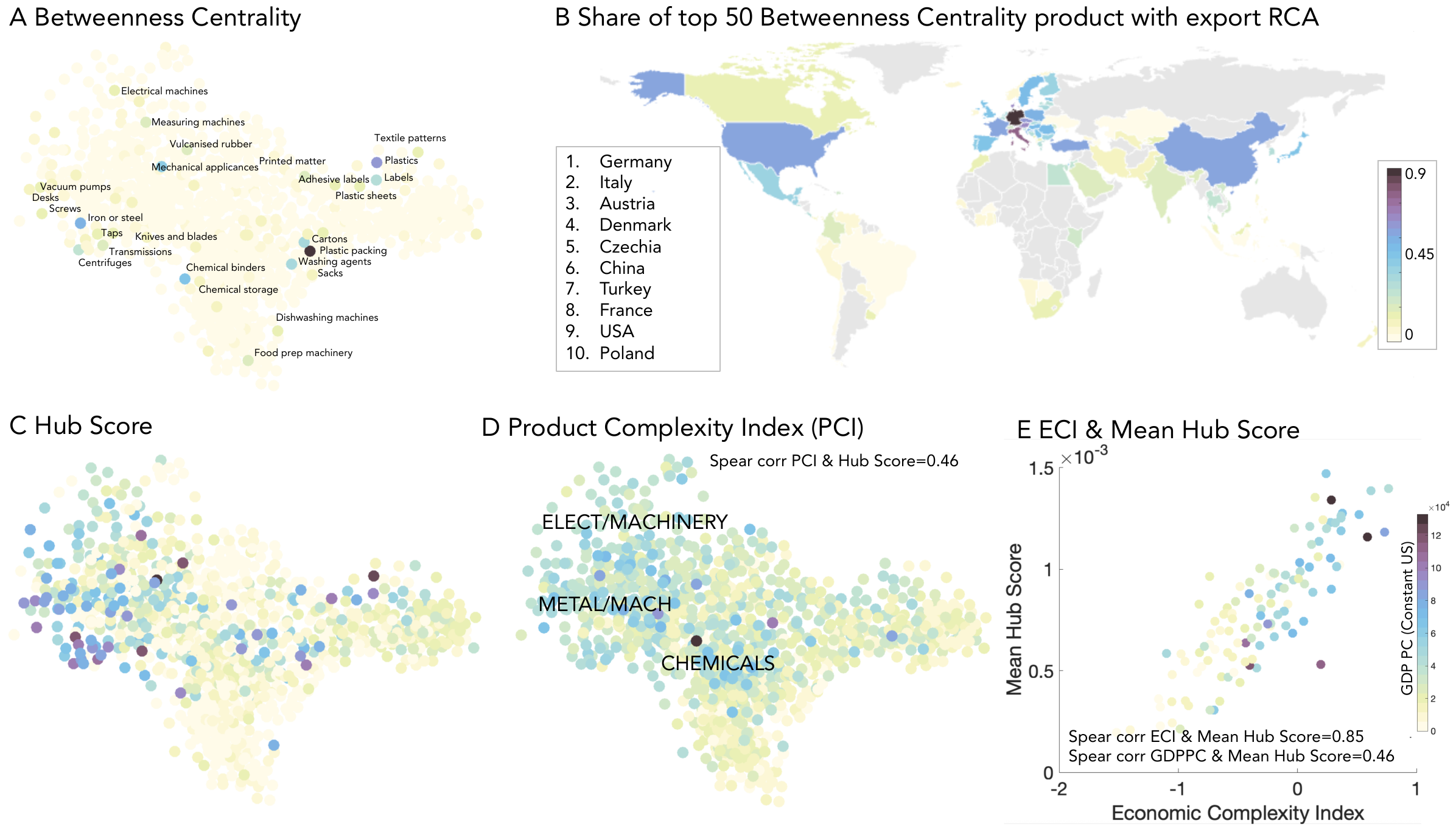}
  \caption{\textbf{European industrial nations and China dominate critical intermediate products, while country-level industrial complexity is correlated with embeddedness in densely connected supply chains.} [A] Various types of machinery, common components such as pumps, blades, taps and other common 'ingredients' are 'chokepoints' with high betweenness centrality (BC). [B] European industrial nations such as Germany, Italy and Austria, as well as China, dominate high BC products. [C] Node ‘hub score’ is similar to eigencentrality in in the sense that it captures dense 'neighbour of neighbour' linkages. [D] The Product Complexity Index (rank) correlates 0.46 with hub score. [E] Economic Complexity Index (ECI) has a high rank correlation of 0.85 with mean hub score. Hence, countries with high industrial complexity tend to export products which are embedded in densely connected global supply chains.}
  \label{fig4}
\end{figure}

\textbf{European industrial nations and China dominate critical intermediate products, while industrial complexity is correlated with embeddedness in densely connected supply chains.} 
Beyond the wider network structure, identifying key nodes in the network, whether they be important chokepoints or critical inputs, is key for a better understanding of leverage, resilience, and systemic risk \cite{farrell2019weaponized}. 

Starting with the simplest node-level metrics, as shown in SI Figure S2, we find that machinery products, including those related to cars and motors, tend to have a large in-degree (they require many inputs), while components, plastics and packaging tend to have high out-degree (i.e., they are inputs to many products). Further, consistent with what we might expect, deploying the UN BEC classification that sorts products into capital, intermediate and consumption goods, we find that capital products are inputs to many products (high out-degree), while both capital and consumption goods have high in-degree – they require many ingredients (SI Figure S2). 

From a network structure perspective, betweenness centrality (BC) measures the extent to which a node acts as a key 'juncture' through which many paths (i.e., supply chains) cross, and is a common proxy for supply chain chokepoints \cite{borgatti2009social, li2020network}. Mathematically, the betweenness centrality of a node is the number of times a path (between any pair of nodes) transits that node. In Figure 4 A, we highlight products with high betweenness centrality, including various types of machinery, common components such as pumps, blades, taps and motors, plastics, packaging, and chemical binders. 
Nations which produce these key inputs have an important role in global supply chains, controlling so-called 'choke points'. To investigate this, we compute the Revealed Comparative Advantage (a measure of export concentration of a country in a product) for each country-product pair from global trade data (Comtrade 2024). In Figure 4 B, we show a global map with countries shaded by their share of top-50 BC products exported with RCA. We find European industrial nations such as Germany, Italy and Austria, as well as China, dominate high BC products. 
 
Beyond being active in critical 'chokepoint' sectors, we would expect that highly industrialised nations are embedded in densely connected global supply chains as modern production depends on large networks of specialized firms, resources, and markets. To investigate this, we ask to what extent are complex goods located in densely connected parts of the network. To capture production complexity, we adopt an approach proposed by \cite{hidalgo2009building} to infer product complexity (PCI) from patterns of geographical production captured in trade data. To measure the localised connectivity and embeddedness of product in global supply chains, we compute the hub score of each node \cite{kleinberg1999hubs}. This is similar to eigencentrality in the sense that it captures 'neighbour of neighbour' linkages, but is suitable for directed networks such as ours. 
We find that the PCI (obtained for 2023 from the Harvard Atlas of Economic Complexity) has a rank correlation of 0.46 with node hub score. At the country level, the Economic Complexity Index (ECI), which is mathematically equivalent to the mean PCI of products exported with RCA, has a high rank correlation of 0.85 with mean hub score. Hence, we conclude that countries with high industrial complexity tend to export products which are highly embedded in densely inter-connected global supply chains. 

\begin{figure}[t!]
    \centering
  \includegraphics[width=1\textwidth]{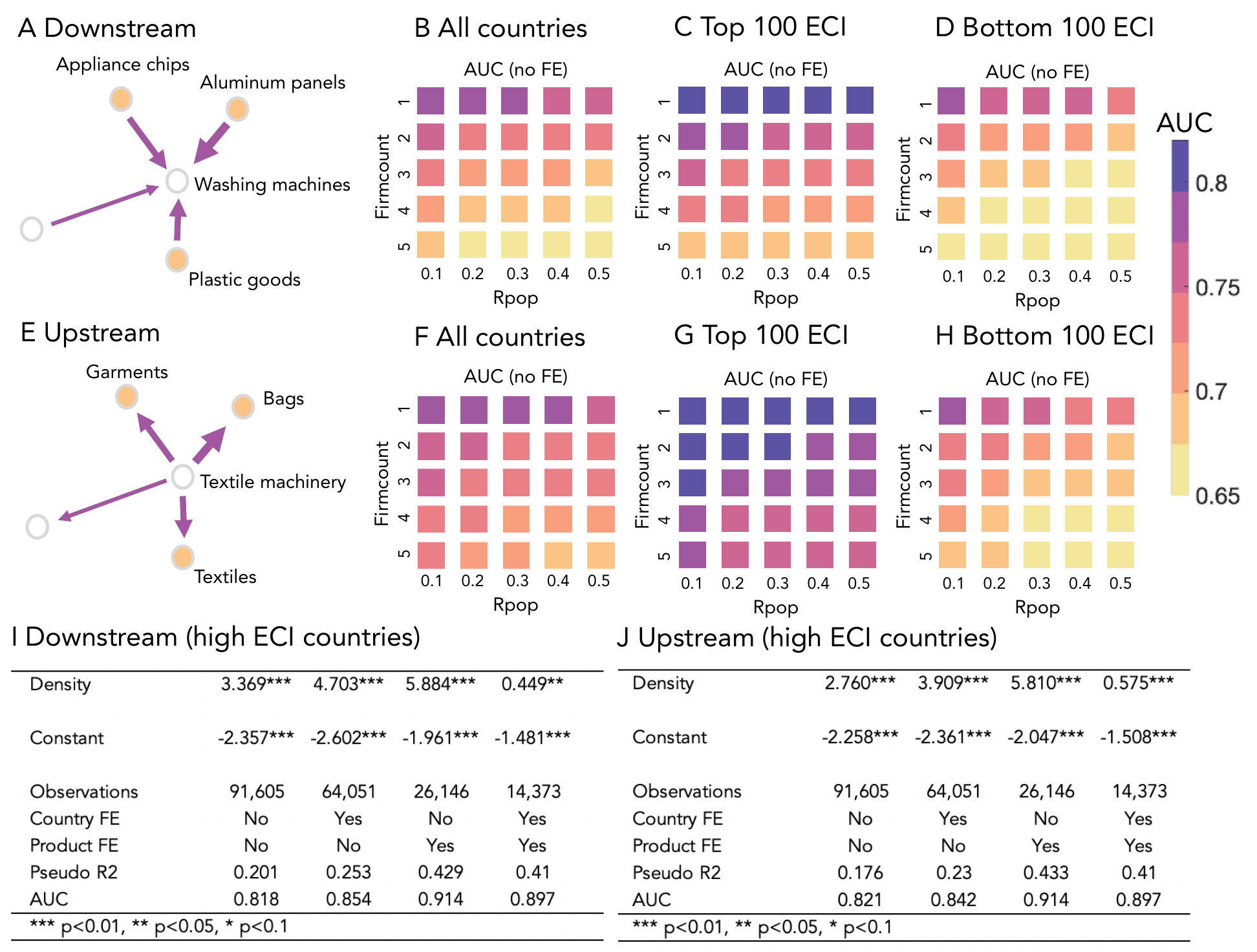}
  \caption{\textbf{Upstream links predict country-product export diversification patterns for countries with higher industrial complexity.} [A/E] Both downstream- and upstream-density captures a country's presence in products that are downstream/upstream of the target product. [B-D] The downstream model (no FE) performs well, in particular for low firmcount (many edges) and top 100 ECI countries. [F-H] The upstream model (no FE) performs consistently very well over a wide parameter space, with highest values for low firmcount (many edges), low increase in Rpop, and top 100 ECI countries. [I/J] We show the full regression table (include FE) for both downstream and upstream models for high ECI countries and parameters firmcount equals 1 and entry Rpop threshold 0.1.}
  \label{fig7}
\end{figure}

\textbf{Upstream links strongly predict country-product export diversification patterns for countries with higher industrial complexity.} 
The network describes the complex structure of global product supply chains. It implicitly also encodes longer-term growth potentials, specifically the probability that a country might diversify into a new 'adjacent' product - whether upstream or downstream - over time. As discussed above, the key logic is that countries tend to already have some of the capabilities and knowhow required for these new products due to their proximity in supply chains. 

To investigate this, we adopt an approach from the Economic Complexity and Evolutionary Economic Geography fields. In this literature, density and 'relatedness' metrics \cite{hidalgo2018principle} capture the propensity for diversification (of a country into a product) based on the proximity (in terms of, for example, supply chain linkages) between what the country currently produces and the new potential product. Specifically, we construct a 'downstream' (and analogous 'upstream') density metric, as illustrated in Figure 5 A/E, for a product $p$ in country $c$ which captures the share of top in-degree neighbouring products of $p$ present in country $c$:  
$$
d_{p,c}=\frac{\sum_{j\in J_p} I(A_{p,j})M_{p,c}}{\sum_{j \in J_p} I(A_{p,j})}
$$
where $J_p$ is the set of top $k=40$ edges (variation in $k$ shown in SI Figure S6) from $p$ to all nodes, and $M_{p,c}=1$ if product $p$ is present in country $c$ (and otherwise 0). We also construct an equivalent 'upstream density' metric which considers upstream neighbouring nodes. 

We aim to predict the appearance of a new product in a country (in 2021) based on the proximity (in global supply chains) of existing products to the new product in a prior base year (2016). Hence, our main explanatory variable density $d_{p,c}$ is computed using product presences in 2016. Similar to \cite{o2021productive}, we perform a standard Probit regression for the probability of a product appearance of the form:
$$
R_{p,c} = \Phi( \alpha +\beta_d d_{p,c}  + \gamma_p + \eta_c)
$$
where $R_{p,c} =1$ if product $p$ was present in country $c$ in 2021 but absent in 2016, $\Phi$ is a normal cumulative distribution function, and $\gamma_p$ and $\eta_c$ are product and country fixed effects respectively. Rpop, which compares a country’s per capita production levels in a product to the world’s overall per capita production of the product, is a threshold used to define product 'presence' or 'absence' (similar to Revealed Comparative Advantage \cite{balassa1965trade}). 

In Figure 5 B-D and F-H, we show the predictive power (the AUC or 'area under the curve') for both models (downsteam and upstream respectively) without fixed effects for three distinct country sets: a full set of 235 countries, the top 100 countries by ECI and the bottom 100 countries by ECI. The grids vary parameters firmcount and Rpop. Figure 5 J and I show the full econometric model (with various fixed effect models) in table format for the subset of 100 high ECI countries with network parameters firmcount equals 1 and entry Rpop threshold 0.1. 

Overall, across variation in model parameters and country subsets, we find that both downstream and upstream linkages are predictive of diversification patterns - but that upstream linkages are significantly more predictive especially for the higher ECI group of countries. This is consistent with an established literature on the role of upstream linkages in diversification processes \cite{hirschman1958strategy, bahar2019export, ndubuisi2021important}. The high ECI set is overall better predicted relative to the full set and low ECI set, which is likely due to the fact that, as we saw above, these countries are more strongly embedded in densely connected global value chains. 

It is also clear that, perhaps counter-intuitively, the predictive power is highest for low firmcount values. Hence, including 'weak links' for which there is relatively lower certainty is informative for the model. We are also better in general at predicting small increases in Rpop over a 5 year period compared to larger jumps in Rpop. However, many model specifications do achieve high AUC for larger values of Rpop, e.g., for Rpop=0.5, the upstream model (for high ECI countries/low firmcount) has AUC close to 0.8.

The AUC values we find are in line with (or exceed) the literature on predicting product entries in countries, and other similar efforts to predict diversification using density metrics derived from a variety of networks capturing 'capability overlap' or linkages between products \cite{hidalgo2007product, o2021productive, hausmann2022implied}.
In the SI we show similar results for a 10 year model (2011-21), and variation in density parameter $k$ (this peaks around $k=40$ for most specifications, which is consistent with previous work \cite{o2021productive, hausmann2022implied}, and used in Figure 5 here).

\begin{figure}[t!]
    \centering
  \includegraphics[width=1\textwidth]{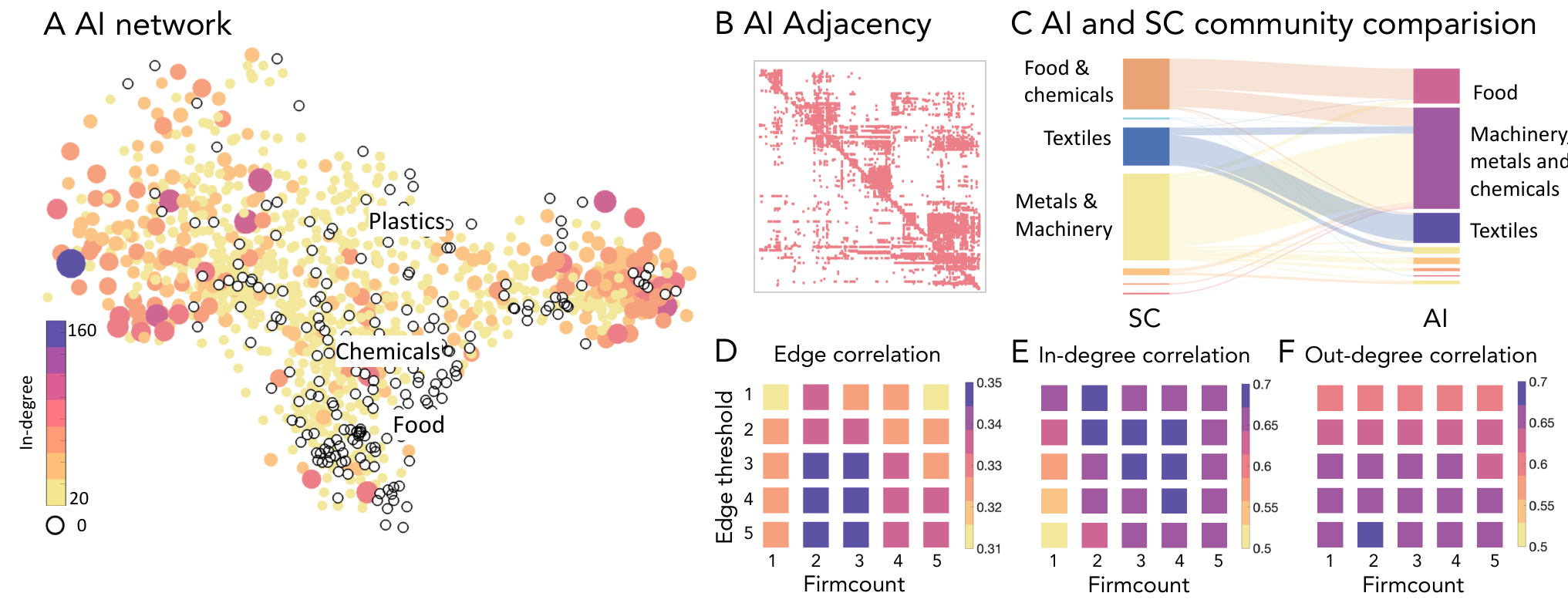}
  \caption{\textbf{The product network correlates with other efforts to construct inter-product links or chains.} [A] The 'AI network' of \cite{fetzer2024ai} is constructed via LLM queries. Nodes with zero in-degree are shown in white. [B] The AI network is much sparser than the SC network and tends to have much fewer edges in sectors such as chemicals, plastics and food. [C] The AI and product network have similar meso-scale structure, with the exception of a much larger and stronger chemicals and food clusters in our network (SC network). [D-F] The correlation of the full edge set of AI network and SC network is approx 0.35, while the in and out-degree edge correlation is significantly higher, peaking at approx 0.68 for in-degree (for low firmcount in the range 2-4).}
  \label{fig5}
\end{figure}

\textbf{The supply chain network correlates with other efforts to reconstruct inter-product supply chain linkages.} 
While there are no 'ground truth' datasets for us to validate against, particularly since we construct a network of importance scores and not value flows, here we compare our network to a recent attempt to construct a similar type of product network using LLM queries \cite{fetzer2024ai}. Specifically, the authors asked an LLM what are the top input products for each product, and constructed a (binary unweighted) network based on its answers.

Figure 6 A and B show that the AI network is in general much sparser than our network (we keep the same node layout as above). Its clear from the distribution of white nodes (those with zero in-degree) that the AI network tends to have much fewer edges in sectors such as chemicals, plastics and food. Overall, our network (the 'SC' network, with parameters set as above) has 4 times as many links as the AI network, but 5.8 times the number of links in the SC food community, 4.5 times the number of links in the SC chemicals community and 3.3 times the number of links in the metals and electric machinery community. We compare communities for the AI and SC network in Figure 6 C. We find that the AI and SC product networks have a similar meso-scale structure, with the exception of much larger and more strongly connected chemicals and food clusters in the SC network. 

Next, we compute a simple correlation of the full edge set of AI network and SC network, finding they are correlated up to approx 0.35 for low values of firmcount (i.e., including uncertain edges). The in- and out-degree edge correlation is significantly higher, peaking at approx 0.68 for in-degree. In general, these correlations peak for low firmcount values in the range 2-4 - i.e., we keep weak edges for a dense network. Hence, while the networks look visually very different, they display non-trivial similarity in both meso structure and in- and out-degree pattern. 

We also compare our SC network to a product network constructed based on NAFTA 'Rules of Origin' \cite{conconi2018final} which aims to identify the origin country of goods in order to determine if tariffs apply. In this case we find a much lower edge correlation of 0.15, a binary in-degree correlation of 0.28 and out-degree correlation of 0.32. Despite the lower correlations, however, the aggregate meso structure retains similarities to the SC (and AI) networks with food, textiles, chemicals and metals/machinery clusters (see SI Figure S4).

Finally, we compare our production network to product-level Eurostat IO data. While the IO data represents monetary flows, and we have importance scores, its nonetheless interesting to compare them. As detailed in SI Section 4.4, this data contains IO flows for 64 CPA product classes, and covers 27 member states and 22 closely affiliated countries. We collapse this matrix (summing across all countries) to a sector level, and compare it to our production network (averaged at 64 product sector level). We find an edge correlation of 0.26-0.42, with maximum correlation value for edge weight>1, and firmcount>2, and weighted in-degree and out-degree correlations of 0.33-0.47 and 0.46-0.7 respectively (across variation in edge weight and firmcount thresholds).

\begin{figure}[t!]
    \centering
  \includegraphics[width=1\textwidth]{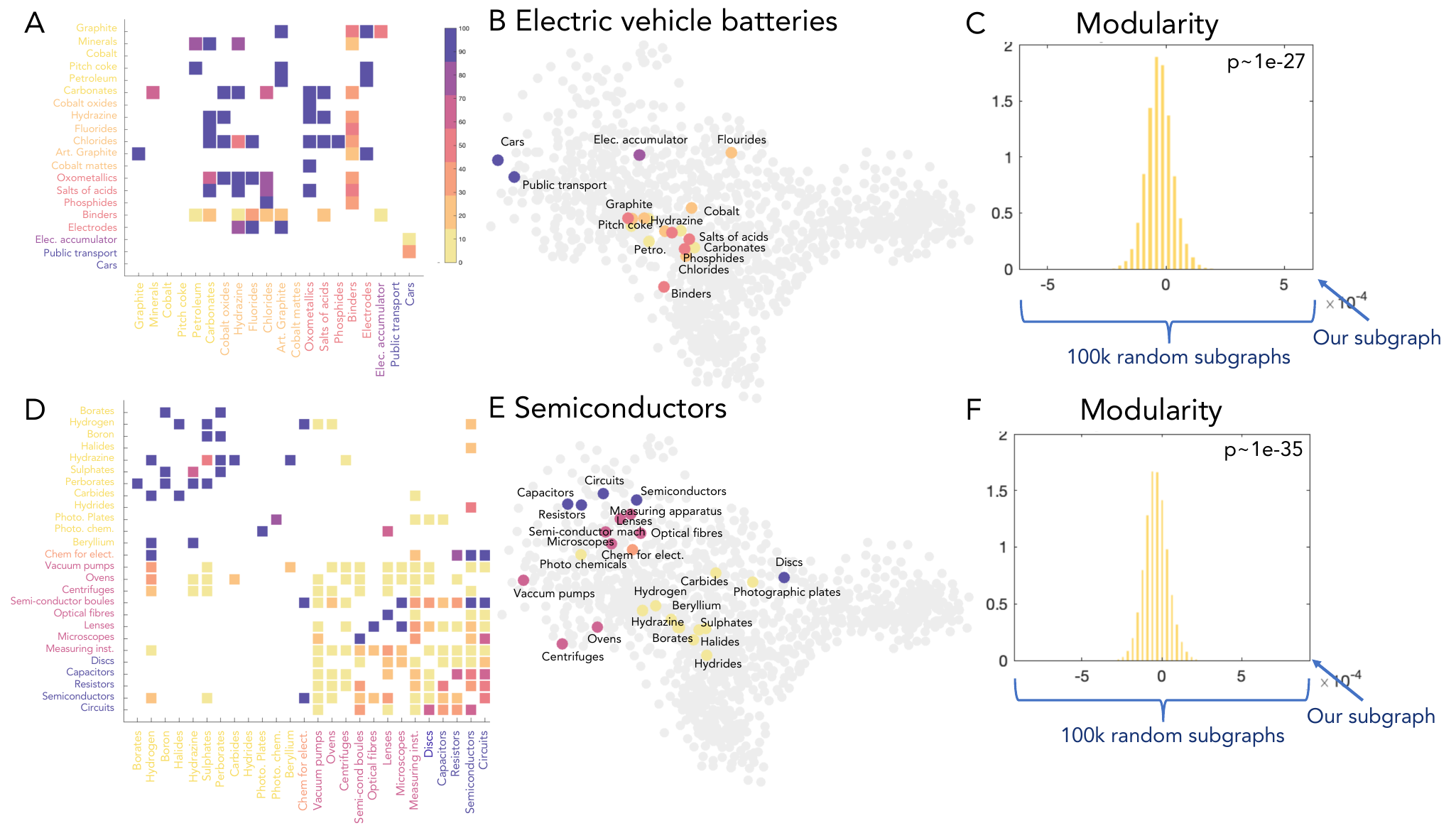}
  \caption{\textbf{The product network has dense connections between products identified in manually constructed single-sector supply chains.} [A/D] We capture a high density of links between products identified in electric vehicle battery and semiconductor supply chains. [B/E] While products in electric vehicle battery are highly clustered in the centre, semi-conductors exhibit two distinct clusters composed of chemicals and machinery. [C/F] To show that we pick up a pattern of dense linkages between these products which is far from random, we simulate 100k random networks with the same number of nodes and compare the modularity score (connectivity) of our subgraph (see A/D) to the random networks. We observe that our subgraphs are in the very extremes of the distributions suggesting they would be extremely unlikely to occur at random.}
  \label{fig6}
\end{figure}

\textbf{The product network has dense connections between products identified in manually constructed single-sector supply chains.}
Up until now, most detailed data on inter-product linkages has come from a large number of detailed manual studies of sector-specific supply chains. As a second quasi-validation step, here we take a selection of these, and investigate to what extent we pick up similar patterns using our method. We focus on two key sectors, critical minerals in electric vehicle batteries \cite{UNCTAD} and semiconductors \cite{haramboure2023vulnerabilities}. 
For each sector, we take the set of products identified in the manual supply chains, and visualise the entries of the adjacency matrix of the subgraph of these products in Figure 7 A/D. In other words, we show the edge pattern that we find in our network for the manually identified product sets. We observe that we capture a high density of links between products identified in electric vehicle battery and semiconductor supply chains. While minerals in electric vehicle batteries are highly clustered in the centre of the network, semi-conductors exhibit two distinct clusters composed of chemicals and machinery products. 

Could these links be random? In other words, if we selected a random set of products, would they also have a similar link density? To test this, we simulate 100k random subgraphs with the same number of nodes (as those identified in the sector supply chain) and compare the modularity score of our subgraph to those of the random subgraphs. The (directed) modularity score \cite{leicht2008community} essentially tells us how modular or connected the subgraph is compared to what would be expected at random under a null model (we use a standard configuration null model which randomly allocates edges while keeping the node degree fixed). We compare the modularity of our network (far right axis) to the distribution of modularity scores for the ensemble of 100k random subgraphs, finding an extremely low probability that our subgraph would be simulated at random (p-value shown).  

\section*{Discussion}

The key contribution, here we develop a highly granular network tool to describe product-level global supply chains. Our production network captures a more complex interdependency structure than the traditionally conceived ‘linear’ or 'tree-like' supply chain. For example, what might traditionally be considered a root node in a supply chain is typically not a root node (i.e., zero or very low in-degree) in our network as activities like mining, for example, require the purchase of high-tech machinery and safety equipment. Hence, we observe a more complex structure in which highly sophisticated ‘final consumption’ goods are purchased by firms at all stages. The centrality metrics we implement - degree centrality, betweenness centrality and the hub score - are suited to this network as they do not implicitly assume a chain-like structure (with root and sink nodes) as do metrics such as closeness centrality \cite{borgatti2009social} or trophic levels \cite{pimm1977number} which have been previously applied to study supply chain networks. 

The network we construct is based on cross-border transactions, and thus captures inputs which are sourced directly by firms in global markets - and not locally purchased inputs which are either, for example, cheap to make locally or expensive to transport. For example, if the water used to produce beer is predominantly purchased locally, we may not capture it as a key input. More broadly, we do not capture: 1) domestically produced and purchased inputs, 2) internationally produced but domestically purchased inputs (e.g., from an importer), or 3) inputs produced in-house by firms. However, given the vast scale of the data, we are likely to still pick up inputs which are only occasionally purchased abroad. Given these constraints, the network is most suited to applications which focus on global supply chain connectivity and, in turn, the metrics we derive, such as critical connective nodes in the network, should be interpreted as chokepoints in cross-border supply chain networks.  

The network structure provides key insights on 'central nodes' and clusters of highly interconnected products. For example, we find that European industrial nations and China dominate critical intermediate products such as metals, common components and tools. Steel, for example, has very high betweenness centrality indicating that it plays a role in many paths (chains) and can be seen as a critical good. This accords with the fact that the US government (and others) moved aggressively to protect domestic production of steel throughout 2025 via steep tariffs on imports of both steel and downstream 'derivatives' which includes products with a high steel content such as bolts and bicycles. 

Building on a substantial and well-established literature on knowledge spillovers (and other positive benefits) between buyer-supply firms (e.g., \cite{dyer2000creating}), we show that the supply chain links captured by the network structure are informative of future product diversification patterns of countries. Consistent with previous findings based on US IO linkages \cite{bahar2019export}, we find evidence for both forward and backward linkages, with a stronger signal for forward (upstream) linkages. In a similar vein to \cite{hidalgo2007product, cainelli2019industrial}, the network could be used as a policy tool to identify 'feasible' opportunities for diversification (in the form of both upstream and downstream products) which could be targets for industrial policy support. This type of policy approach is used, for example, in 'Smart Specialisation' efforts at a European level in which regions aim to specialise and diversify in related sectors \cite{balland2019smart}. 

Beyond those illustrated here, there are a wide range of potential research and policy uses for this type of tool. While we don’t seek to explicitly model particular shocks here, modelling shocks or cascades on industry and industry-country networks is well established \cite{acemoglu2012network, carvalho2014micro, piccardi2024patterns}. For example, one could implement a cascade-type model in which nodes (or edges) are removed and propagating shocks are simulated \cite{battiston2012debtrank, hearnshaw2013complex} and, akin to \cite{chakraborty2024inequality}, extend this approach to investigate inter-country shock propagation processes and vulnerabilities. More broadly, the network encodes complex economic and geo-political inter-dependencies between countries, and can be used to study policies such as tariffs and subsidies which aim to exert pressure and influence \cite{clayton2023framework}. 

In terms of wider applications, there is a growing literature on 'embodied carbon' as carbon from input products is incorporated to final goods \cite{sato2014embodied}. Not enough is known, however, about so-called Scope 3 emissions - or those transmitted via supply chains - despite the fact they are thought to be growing faster than other types of emissions \cite{hertwich2018growing}. Hence our network, which captures global production paths, could be used to track cross-border 'product life-cycle' emissions as products are manufactured and then exported to be used in other manufacturing processes \cite{wang2011green}. Understanding how carbon 'relocates' in global supply chains is central to the successful implementation of Carbon Border Adjustment Mechanism (CBAM) policies, a landmark effort designed to prevent 'carbon leakage' across borders (i.e., when production shifts to regions with less strict emissions rules). Already in force in the EU, the UK aims to bring in its own version of CBAM in 2027. 

There are some further limitations to our approach, as well as potential sources of noise. Firstly, in order to build our network, we aggregate data over 3 years based on advice from the supplier and our analysis of the raw firm-interaction data that suggests the data is most complete for these latter years. This approach, however, does not allow for temporal analysis. Future work may investigate the extent to which disaggregating into distinct time periods and recomputing the network for each is feasible. Secondly, there is no doubt some noise in our data due to the uneven composition of the firm-level data. For example, while we have good coverage of firms across developed countries, we by no means have the 'universe' of all global firms in our dataset, and some countries/sectors will be more represented than others. This issue is less problematic for our tool than say country-specific or firm-level analysis, but nonetheless it will introduce noise which we cannot fully mitigate against. 

Third, there may be other sources of noise, such as short term changes in purchasing patterns. For example, we noticed an unusual input to semiconductors - washing machines. It turned out, however, that due to a huge shortage in input components in 2022/23, semiconductor firms were actually buying washing machines to strip them of their chips \cite{washing}. Another source of noise is likely to be multi-product firms. While we have taken efforts to reduce this by omitting firms that sell products in more than five HS Sections, and our overall approach based on common purchases by all firms that make a product, there will no doubt be some noise remaining due to this issue. Fourth, there may be idiosyncratic differences between supply chain relationships across regions that we aggregate over by computing a network using all firm transactions. For example, battery supply chains may look different in the US vs Asia. In future work we may reconstruct the network using data for specific regions to further investigate this issue. Finally, we present the network at HS 4-digit level. It may be possible in future work to construct the network at 6-digit level, but due to an exponential increase in the number of edges to estimate, we have much fewer number of firms (firmcount parameter) for each linkage and fewer sources of external data to compare the network structure to. Again, we leave this to future work. 

\section*{Materials and Methods}

\subsection*{Processing of underlying firm data}

Our GSCIP data includes firm-level transactions during the time period between January 2021 and December 2023. We take the following processing steps: 
\begin{itemize}
\item We assign all firms to an 'owner firm' based on ownership information available to us (see SI).
\item We aggregate all firms to an 'owner-country' level (see SI). In other words, we have Coca Cola-Japan and Coca Cola-Belgium. 
\end{itemize}

\subsection*{Multi-product firms}

A concern about our approach might be that it could pick up spurious links due to the presence of firms in our dataset which make many products (and hence we cannot infer specific inputs to products from their purchases). Our methodology is designed to reduce this issue since we look at the purchases of not just one firm but all firms that make a specific product. To mitigate this potential issue further, we drop all firms that export products in five or more HS Sections (e.g., this would exclude wholesalers and multi-nationals with large portfolios of diverse products such as Proctor and Gamble). As expanded on in SI Section \ref{SIwholesale}, these constitute about 4\% of all firms in our dataset and include mainly manufacturing and finance-related firms (e.g., holding/parent companies).   

In order to further reduce noise, in addition to the firmcount parameter as mentioned above, we also drop any links from product $i$ to product $j$ in which the total value of the transactions of the firms in set $S^i_j$ is less than \$1000. 

\subsection*{Network analysis}

\textbf{Community detection} We use the Stability algorithm of \cite{delvenne2010stability} to detect communities in the SC and AI networks. The algorithm detects partitions on a range of 'scales' corresponding to the 'time' spent by a random walker exploring the network. The lower the time, the smaller (but more numerous) the communities detected. The longer the time, the larger (and fewer) the communities. We use the 'full' (non-linearised) specification for a directed Laplacian, with 100 iterations. In the SI we show the number of communities, variation of information across the iterations (a low value tells us we find similar partitions for different random seeds), and the stability score for our set of partitions corresponding to a range of timescales. 

\textbf{Modularity score} We use the Modularity score to compare the connectivity of the subgraph induced by each of our sector-specific supply chains (EV batteries and semi-conductors) to 100k randomly generated subgraphs of the same size. Modularity \cite{newman2006modularity} compares the connectivity of a community to what would be expected under a random null model - in this case a network in which the node degree is preserved but the edges randomised. 

Effectively, we decompose the directed Modularity score proposed by \cite{leicht2008community}, which aims to evaluate the connectivity of a set of communities in a network, and use it instead to evaluate the connectivity of a single subgraph (i.e., a single community). This is mathematically coherent as Modularity is a sum over the connectivity within communities. Specifically, we compute:
$$
M_G=\frac{1}{m}\Large(\sum_{i,j \in G} X_{i,j}-\frac{Out_i  In_j}{m}\Large)
$$
for node pairs $i,j$ in subgraph $G$, $X=A>0$ (unweighted adjacency matrix), $m$ is the number of edges in the full network, $Out_i$ is the out-degree of node $i$ and $In_j$ is the in-degree of node $j$.

\subsection*{Econometric analysis}

To capture product entries, we wish to compute a measure of a countries' concentration in a product. The traditional approach is Revealed Comparative Advantage (RCA), proposed by \cite{balassa1965trade}, captures the proportion of the country's exports in a product divided by the global proportion of exports in a product. We use a related approach, denoted Rpop, which helps remove noise due to fluctuations in commodity prices, seasonal employment or currency exchange rates \cite{hidalgo2021economic}. Specifically, it compares a country’s per capita production levels in a product to the world’s overall per capita production of the product:
$$
Rpop_{p,c}=\frac{E_{p,c}/pop_{c}}{E_{p}/pop}
$$
where $E_{p,c}$ is the total exports of country $c$ in product $p$, $E_{p}$ is global exports of product $p$, $pop_{c}$ is the total population of country $c$ and $pop$ is the global total population. 

We conduct the full econometric analysis for a set of 235 countries. We drop products not in the main connected component of the network, which is dependent on the values of parameters firmcount and edge threshold, and 28 products with global trade in 2021 less than \$10m. For the tables shown in Figure 5 I/J, with both firmcount and edge threshold equal to 1, we have $n=1198$.
Of a possible 100x1198 observations in Figure 5 I (high ECI country set), 76\% were absent ($Rpop<0.05$) in 2016. Of these, 3.6\% of products appeared by 2021 (with $Rpop>0.1$). 

\newpage
\renewcommand\linenumberfont{\color{white}}

\bibliographystyle{sciencemag}
\bibliography{references}

\section*{Acknowledgements} 

We would like to thank Christian Diem, Tobias Reisch, Adnan Khan and Fergus Cumming for their useful input and comments, and Shaun Hoang and Sukankana Chakraborty for their research assistance. The data and computing infrastructure was provided by the the UK Government Global Supply Chain Intelligence Programme (GSCIP). 

\textbf{Funding:} NOC would like to acknowledge funding from the 2024-2025 UKRI Policy Fellowship programme (ES/Y004817/1\_583688).

\textbf{Author contributions: } \\
Conceptualization: NOC, DTH \\
Methodology: NOC, BRB, TS, DTH \\
Investigation: NOC, BRB, TS \\
Visualization: NOC \\ 
Supervision: NOC, BRB, DTH \\
Writing-original draft: NOC \\
Writing-review \& editing: NOC, BRB, TS, DTH \\

\textbf{Competing interests:} Any views expressed are solely those of the author(s) and so cannot be taken to represent those of the Foreign Commonwealth and Development Office (FCDO), or to state FCDO Policy. This paper should therefore not be reported as representing the views of the FCDO, its Ministers or His Majesty's Government.

\textbf{Data and materials availability: } The underlying firm-level data was made available to the research team via the UK Government Global Supply Chain Intelligence Programme (GSCIP). This sensitive commercial data is held under license is, unfortunately, not available to external researchers.  

\newpage

\renewcommand{\thefigure}{S\arabic{figure}}
\setcounter{figure}{0}
\renewcommand{\thetable}{S\arabic{table}}
\setcounter{table}{0}
\renewcommand*{\thesection}{\arabic{section}}
\setcounter{section}{0}

\section*{}

\begin{center}
\LARGE{Supplementary Materials for  
\vspace{0.7cm}
\newline
\textbf{Deciphering the global production network from cross-border firm transactions}} 
\vspace{0.7cm}
\newline
\vspace{0.5cm}
\normalsize{Neave O'Clery* et al. 

*Corresponding author. Email: n.oclery@ucl.ac.uk}
\end{center}



\section{Data sources and coverage}

This paper relies on data accessed through the UK HM Government's Global Supply Chain Intelligence Programme (GSCIP). This contains firm-to-firm cross-border transactions, derived from customs and shipping data, which was compiled by a commercial provider. Further information on firms in the transaction dataset has been gathered by GSCIP from business registries, tax filings, company websites, and other sources. 

The dataset includes 'primary' customs data from a set of available country sources. All other countries are represented in the dataset as both customers and sellers of these primary countries. The exact set of primary countries included within the GSCIP programme is suppressed for reasons of sensitivity. We do, however, provide an overview of country-level coverage relative to total trade volumes in Figure 1. 

While GSCIP is an ongoing programme, we use data from the period from January 2021 to December 2023, accessed in early 2025.

\section{Processing of underlying firm data}

Due to the noisy and complex nature of global supply chains and firm-level transaction data, we completed additional cleaning and processing of the raw data as detailed below. 

\subsection{HS 2022}

As some transactions in our data use the previous 2017 Harmonised System of product classifications we convert these to the 2022 Harmonised System so that we have a consistent set. This was necessary as our data includes 2021 when the HS 2017 convention was still in place, and there were also cases where firms in later years had not updated their codes to the newer standard when completing customs declarations. We followed concordance tables produced by the Statistical Division of the UN \cite{comtrade} for the conversion of products between the different harmonised systems. 

\subsection{Firm ownership}

Using GSCIP ownership data, we complete a process to agglomerate all firms to an 'owner-country' level. 

The reconciliation of subsidiaries within country is important as the underlying data has instances whereby a subsidiary of a multi-national imports a product which is then used by a separate subsidiary of the same multi-national. In this case the imported goods are transferred between the subsidiaries domestically (under the same owner umbrella), but we do not capture these transfers as our data contains only cross-border trade. For example, Cadberrys may import cocoa beans using one subsidiary and produce a chocolate bar using another subsidiary. 

Transacting firms are reported at various levels within a company ownership structure; some may be wholly owned, while others might be a subsidiary of a multinational (or a subsidiary of a subsidiary … of a multinational). For each firm-entity we climb the ownership hierarchical structure, provided by GSCIP, until we get to the highest level of ownership up to the point of operation (as we do not want to classify distinct firms i.e. Heinz and Duracell into a larger non-distinct owner such as Berkshire Hathaway). To do this we stop climbing the ownership hierarchy when the next firm up is a financial entity or corporation. Approximately 5\% of firms are assigned a parent firm in this way. 

 

\subsection{Multi-product firms \label{SIwholesale}}

We filter out firms that export in 5 or more HS-sections – removing the most prolific wholesalers from our transaction set. As shown in Table \ref{TabSI1} A, this filtered set made up 4.5\% of the firm set. We experimented with the wholesaler threshold, settling with a relatively lenient threshold of 5 or more HS-sections. 

Beyond HS product codes, we can also look at the NACE classification of the firms. With this we can better understand the types of firms we are filtering out under each of the wholesaler thresholds. The wholesaler filer at all thresholds is most punitive to firms classified as 'NACE C – Manufacturing'. A more punitive wholesaler filter removes more manufacturers. In three of four cases (with two or more HS-sections being the exception), 'NACE K – Financial and Insurance Activities' is the second most attributed wholesaler sector. These appear to be parent/holding companies – where their subsidiaries (and attributed transactions) are varied across sectors.

Finally, we note that while we have NACE categorisation 'NACE G - Wholesale and retail trade' for a large proportion of the firm set, we find that this does not cover the multi-product/sector firms we seek to drop from our dataset to construct the product network (as they tend to be concentrated in other categories such as manufacturing and finance as above). Of firms designated a wholesaler (NACE G), we remove X\% (i.e., those who sell in more than 5 HS2 Sections).

\section{Firm-level transaction network}

Here we present descriptive statistics on the underlying inter-firm transaction data. 

\subsection{Degree statistics}

Similar to other work to understand the structure of firm-level production networks \cite{bacilieri2023firm}, we analyse the in- and out-degree of our firm population. 

The 1 billion transactions in our data correspond to 47 million firm-to-firm links. We consider this data to form a network in which nodes are firms and weighted edges correspond to the sum of the value of transactions between these firms within our time-period. The network is highly heterogeneous: the majority of firms (63\%) maintain trading relationships with just one partner, while a minority engage with a large number of partners. We find that the maximum out-degree (number of export partners for a single firm) reaches 351,000, and the maximum in-degree (number of import partners) is 113,000. Notably, 1\% of all firms have more than 50 trading partners, while only 0.01\% have over 1,000 partners.

Within our full dataset of 47 million firms, we observe that 43\% are exclusively exporters, 48\% are exclusively importers, and only 9\% both import and export. This latter set includes a core set of 1.8 million firms which are use to construct our production network (the remainder - exporter- and importer-only firms - are origins or destinations of transactions from this core set).  

\subsection{Firm types by geography}

There are 8,895,000 firms (43\% of all firms in the dataset) that export only (i.e., without any recorded imports). This pattern is especially strong in China (90.1\% sender-only), Taiwan (81.1\%), Switzerland (80.0\%), Hong Kong (75.3\%), Japan (72.8\%), and South Korea (70.6\%). 

There are 9,771,000 nodes that only import (have no recorded exports). The representation of import-only nodes is most notable in Ecuador (88.8\%), Peru (87\%), the Philippines (81.5\%), Canada (74.7\%), and Ukraine (67.6\%). 

There are 1,803,000 firms (9\% of all firms) that export and import within our dataset. These firms are especially prevalent in Vietnam (21\%), Turkey (18\%), the United Kingdom (17\%), India (16\%) and Germany (16\%).   

\subsection{Correlation with trade data}

We compare correlate total exports and imports with aggregate out- and in-transaction values at the country, country-HS2 product, and HS2 product level. The trade data is derived from UN Comtrade, and aggregated over 2021-23 to match our transaction data. We note that, while there is over 250 geographies represented in the GSCIP dataset, Comtrade includes just 136 reporting countries. Those missing include mainly very small and low income countries, but also Russia, Taiwan, Argentina and Bangladesh.

We find very high correlations (>0.97) for product imports and exports, i.e., for inflows and outflows corresponding to the sum over all transactions in each HS2 product category (96 categories). We also find good correlations at the country level (132 countries) in the range 0.62-0.65, with slightly higher correlations for exports vs imports. 
Splitting countries into two sets by gdp per capita (using data from the World Bank), we find significantly higher correlations (0.86-0.89) for less developed countries relative to wealthier countries (0.67-0.69). When we drop outliers in the data (those with residuals greater than 3$\sigma$), we increase the correlations by about 0.1 for both sets - but poorer set remains better correlated (0.94-0.95). Outliers include countries in Southeast Asia such as Thailand and Indonesia.

Finally, we compute a more dis-aggregate correlation of country x HS2 observations. Correlations remain high and, consistent with above, we find better correlations for exports vs imports, and for poorer countries relative to wealthier ones. 


\begin{center}
\begin{tabular}{|l|c|c|c|c|}
\hline
Observations & No. obs. & Correlation & Filt. obs. & Filt. corr. \\
\hline
Country exports      & 132 & 0.65 & 128 & 0.68 \\
Country imports      & 132 & 0.62 & 128 & 0.44 \\
Richer country exports & 66 & 0.69 & 63 & 0.77 \\ 
Richer country imports & 66 & 0.67 & 64 & 0.87 \\
Poorer country exports & 66 & 0.86 & 64 & 0.94 \\ 
Poorer country imports & 66 & 0.89 & 64 & 0.95 \\
\hline
HS2 exports          & 96 & 0.97 & 93 & 0.99 \\ 
HS2 imports          & 96 & 0.99 & 93 & 0.99 \\
\hline
Country-HS2 exports      & 11264 & 0.67 & 11161 & 0.67  \\ 
Country-HS2 imports      & 12319 & 0.58 & 12218 & 0.59 \\
Richer Country-HS2 exports & 6005  & 0.73 & 5942 & 0.79 \\
Richer Country-HS2 imports & 6223  & 0.61 &  6096 & 0.65 \\
Poorer Country-HS2 exports & 5259  & 0.80 & 5225 & 0.93 \\
Poorer Country-HS2 imports & 6096  & 0.91 & 6052 & 0.87 \\
\hline
\end{tabular}
\end{center}




\section{Additional analysis}

\subsection{Alternative specifications} 

For our main analysis, we set stringent conditions to identify firms that both produce and purchase specific products:
\begin{itemize}
\item A firm $k$ produces product $j$ if they sell a larger monetary value of $j$ than the average firm that sells product $j$.
\item A firm $k$ buys product $i$ if they purchase a larger monetary value of $i$ than the average firm that buys product $i$.
\end{itemize}

We also produced alternative networks with variations to these criteria, including:
\begin{enumerate}
    \item filtering the network to just include firms buying more of product $i$ than the average firm, without any conditions on selling product $j$. 
    \item filtering for just firms selling more of product $i$ than the average firm, without any conditions on buying product $i$.  
    \item Including all firms selling product $j$ and buying product $i$.
\end{enumerate}

\begin{figure}[t!]
  \centering
  \includegraphics[width=1\textwidth]{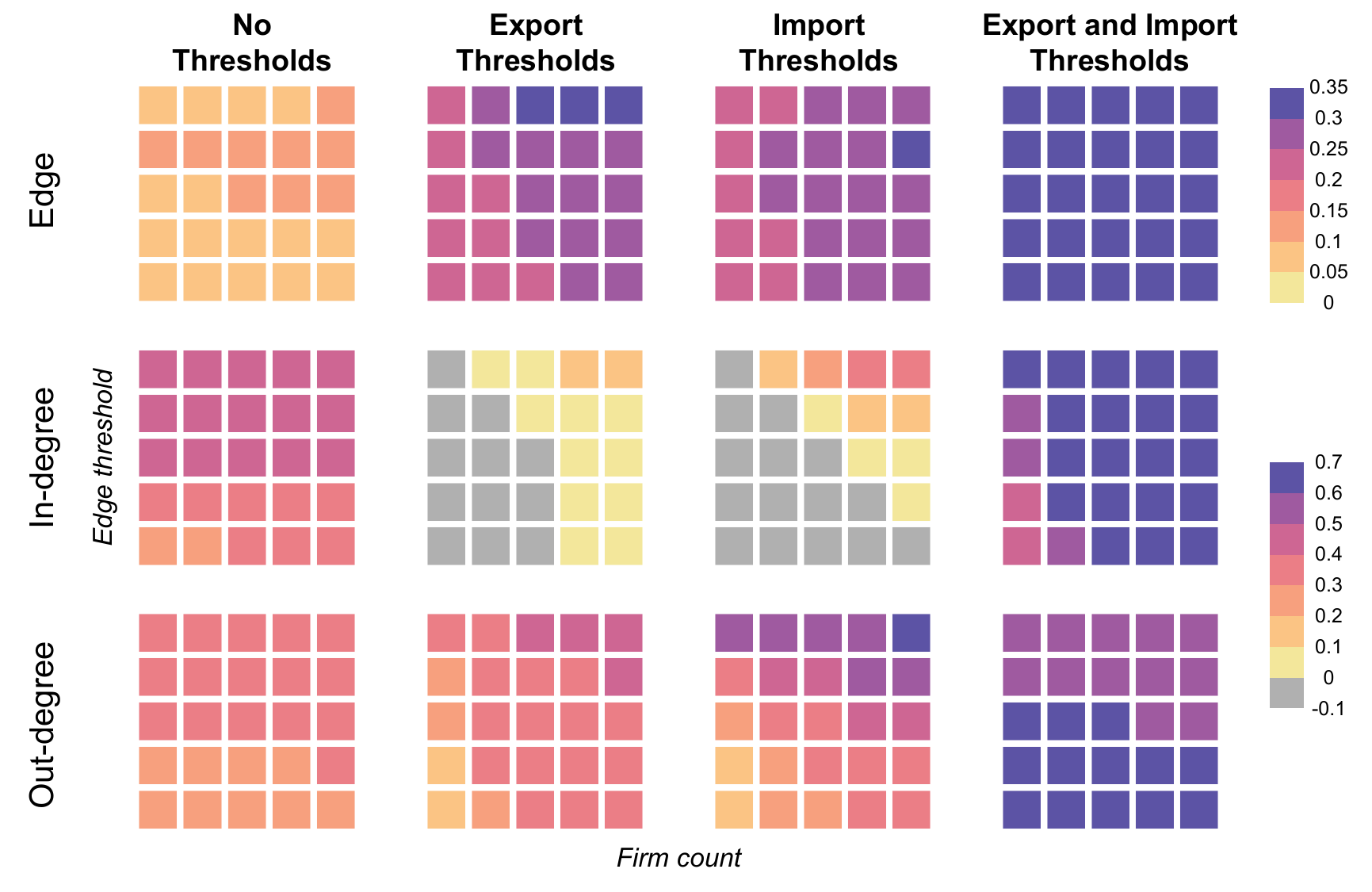}
  \caption{Variations in the underlying formula that forms the product network: edge, in and out-degree correlations with the AI network. Figures show variation for setting Firm count and edge threshold at 1, 2, 3, 4 and 5. The correlation of the full edge set of AI network and SC network is highest with thresholds for both export and import.}
  \label{FormulaSIfigure}
\end{figure}

We replicate Figure 7 in the main text which displays edge, in-, and out-degree correlation with the 'AI network' in Figure \ref{FormulaSIfigure}. We find that the conditions we apply in the main analysis result in stronger correlations with the AI network, regardless of thresholds on edge or firm count. 

\subsection{Community detection} 

We present the number of communities, the Stability coefficient and the Variation of Information for the SC and AI networks for each partition in Figure \ref{SIfig2}. As the partition number increases, the algorithm runs for a longer 'time', and larger (but typically fewer) communities are detected. 

At each scale (partition number), the algorithm seeks to minimise the Stability coefficient and hence the communities are better defined for lower values. In addition, a low value of the Variation of Information tells us we find similar partitions across a 1000 random seeds. We observe that both the SC and AI network have very well defined communities at large partition numbers (low Stability and low Variation of Information), and hence we focus on communities in this region. In Figure \ref{SIfig2} C-E, we visualise partitions 20, 25 and 30. 

In Figure \ref{SIfig3} we show two partitions of the product network constructed based on NAFTA 'Rules of Origin' \cite{conconi2018final} which exhibit similarities to the SC and AI networks.

\subsection{Additional econometric regressions} 

In Figure \ref{SIfig1a} we replicate Figure 5 B-D and F-H for 10 year growth regressions (2011-2021), finding very similar results. 
In Figure \ref{SIfig1a}, we vary the 'number of neighbours' parameter in the density metric for high/low ECI countries, for both downstream and upstream models. Overall there is a small but perceptible peak in the AUF (no FE) around $k=40$ for most models, and so we use this value in the main paper. This is consistent with previous work \cite{o2021productive} which found a peak of around $k=50$ in similar contexts. 

\subsection{Eurostat IO comparison} 

We compare our production network to the Eurostat product-by-product input output data (2022) which covers 27 member states and 22 closely affiliated countries. As our production network is built using cross-border transactions, we drop within-country flows from the IO matrix and average the cross-country flows, resulting in a 64x64 matrix. 

Products in the Eurostat IO table are presented with the CPA 2.1 product codes. We map products in our production network from HS4 to CPA 2.1 by first mapping products from HS4 to CPA 2.2 by means of a correspondence table, and by then mapping CPA 2.2 to 2.1 (using another correspondence table). There is no direct mapping from HS4 to CPA 2.1. 

The HS-to-CPA 2.2 correspondence table is provided at HS6 level and in some cases there are one-to-many HS4-CPA mappings. To handle this we weight the mappings when aggregating from HS6 to HS4. For instance, if four HS6 products can be aggregated to the same HS4 product code, with one out of the four mapping to a different CPA code, then we split the weight of the mappings as 0.25 and 0.75. Once 1228 HS4 products have been mapped to the 64 CPA codes in the IO table, we aggregate our production network to a network with 64 nodes by taking a weighted average of the edges. 

Correlations are obtained for different edge score and firm count thresholds (varied between 1 and 10) as shown in Figure \ref{SIfigIO}. Overall, we find a correlation in the range of 0.26-0.42, with maximum for edge weight>1, and firm count>2. We also correlate the in-degree and out-degree sequences. Since the IO network is virtually complete (i.e., very few zero edges), we compute only the weighted degree sequences. For the aggregated production network, we compute both the binary (unweighted) and weighted (mean in- and out-scores) degree sequences. We find correlations in the range 0.37-0.49 (binary in-degree), 0.23-0.39 (binary out-degree), 0.33-0.47 (weighted in-degree), and 0.46-0.7 (weighted out-degree). 

Finally, while the binary AI network does not represent monetary flows, we correlate the in and out-degree sequences with those of the weighted Eurostat IO network to get 0.31 and 0.33 respectively. 

\newpage
\begin{table}[t!]
  \centering
  \includegraphics[width=1\textwidth]{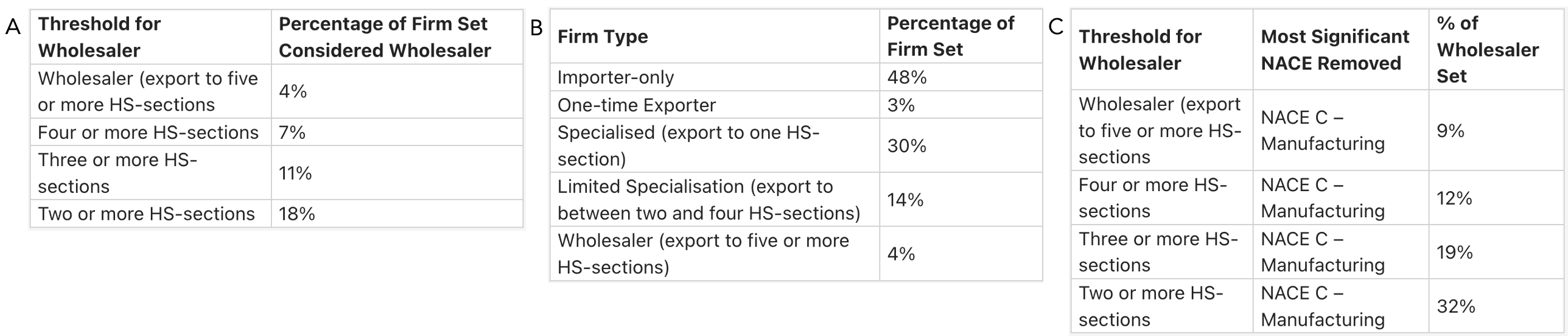}
  \caption{Variations in the underlying formula that forms the product network: edge, in and out-degree correlations with the AI network. Figures show variation for setting Firm count and edge threshold at 1, 2, 3, 4 and 5. The correlation of the full edge set of AI network and SC network is highest with thresholds for both export and import.}
  \label{TabSI1}
\end{table}

\newpage 
\begin{figure}[t!]
  \centering
  \includegraphics[width=1\textwidth]{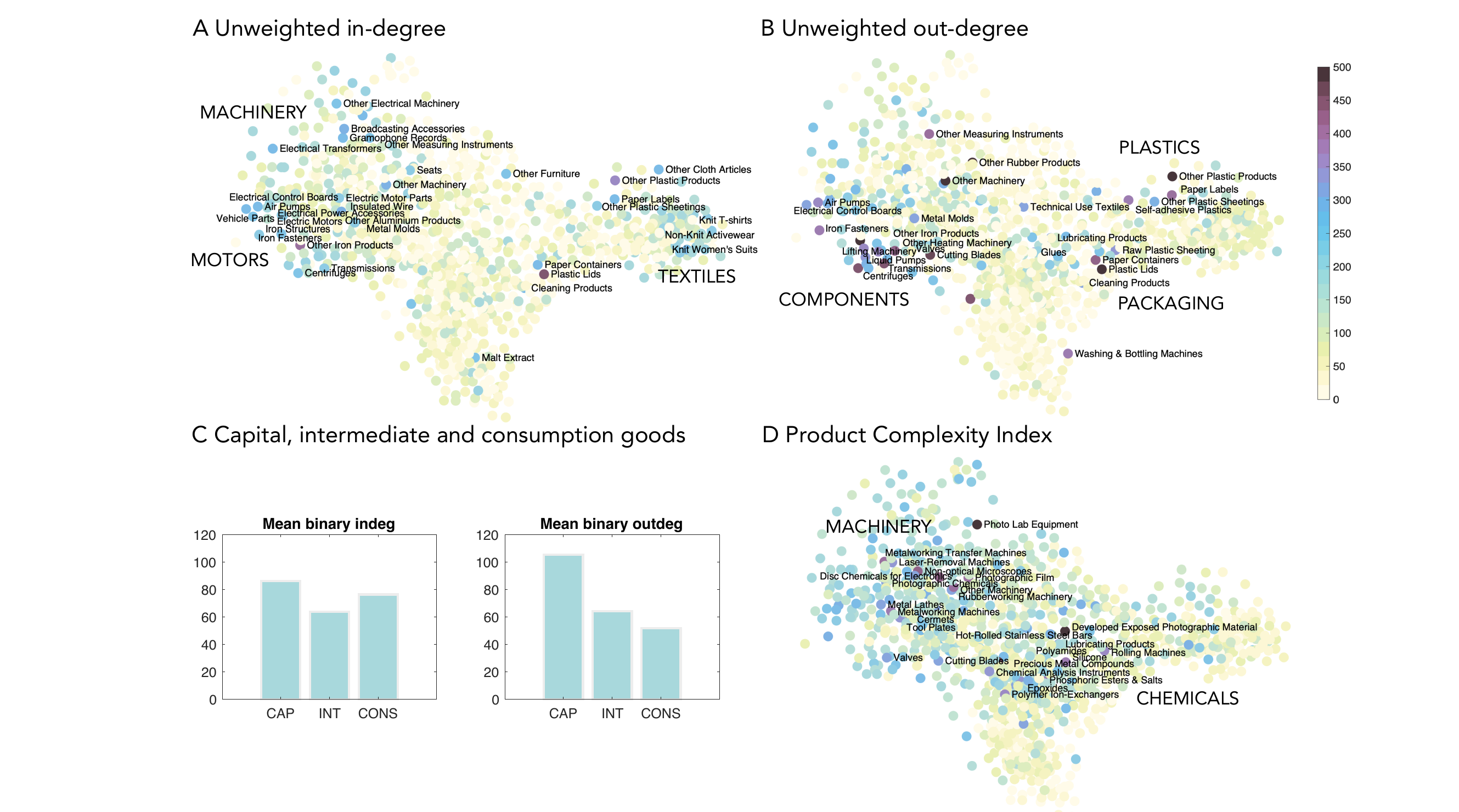}
  \caption{[A-B] In and Out Degree. [C] Capital, intermediate and consumption goods (UN definition).}
  \label{SIfig0}
\end{figure}

\newpage
\begin{figure}[t!]
  \centering
  \includegraphics[width=1\textwidth]{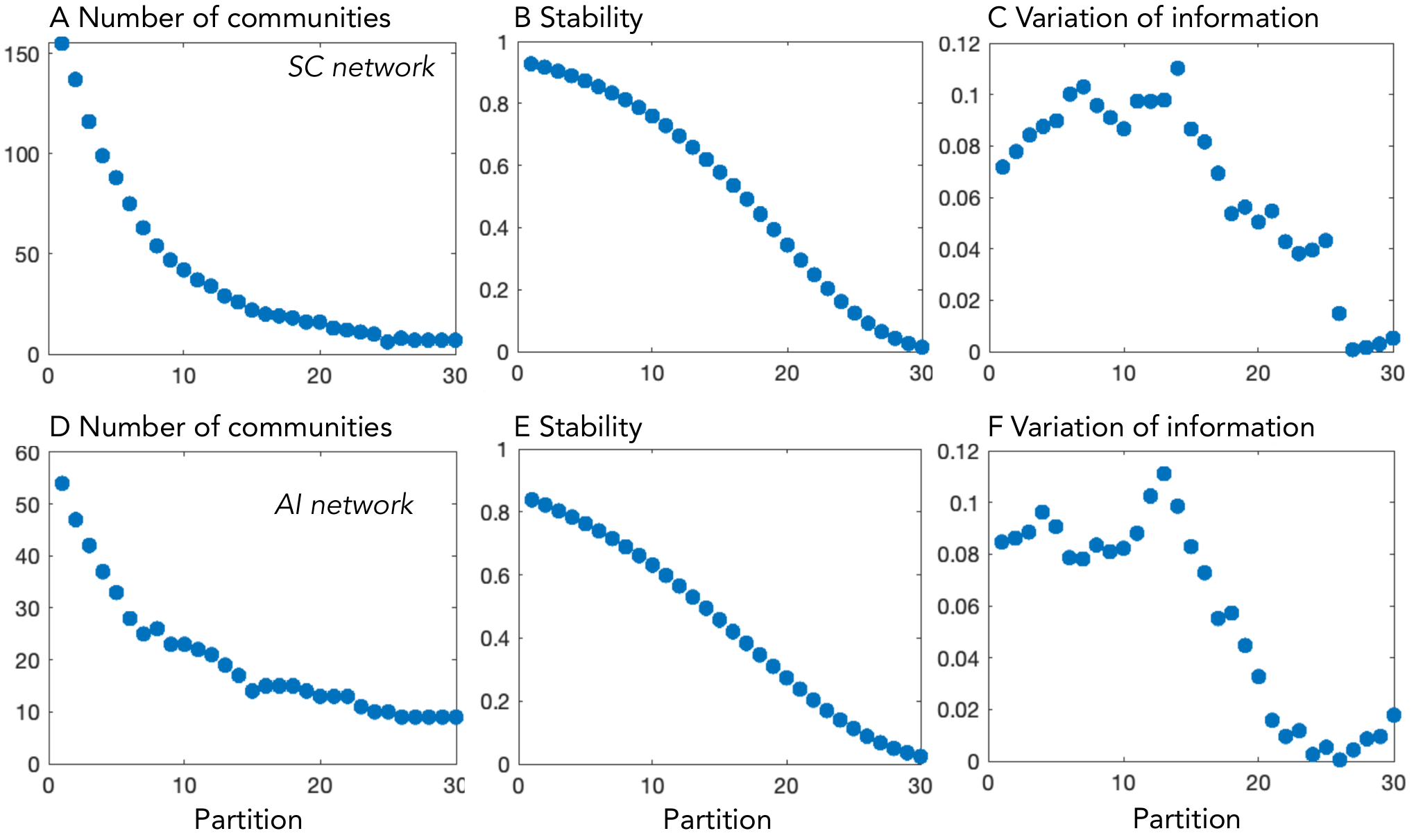}
  \caption{Community detection statistics: Number of communities, stability and variation of information for SC and AI networks.}
  \label{SIfig2}
\end{figure}

\newpage
\begin{figure}[t!]
  \centering
  \includegraphics[width=1\textwidth]{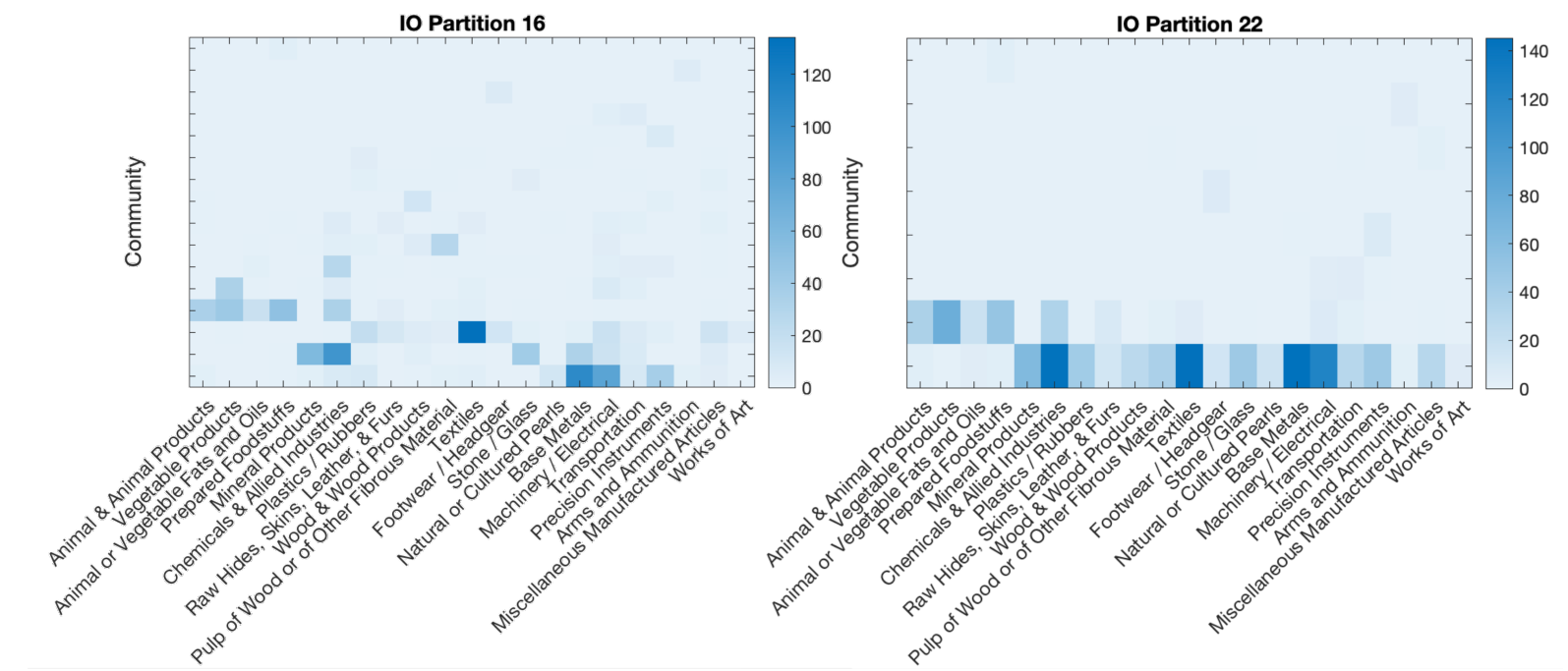}
  \caption{IO communities: Two partitions of the NAFTA IO network, which exhibit similarities to the SC and AI networks.}
  \label{SIfig3}
\end{figure}

\newpage
\begin{figure}[t!]
  \centering
  \includegraphics[width=0.8\textwidth]{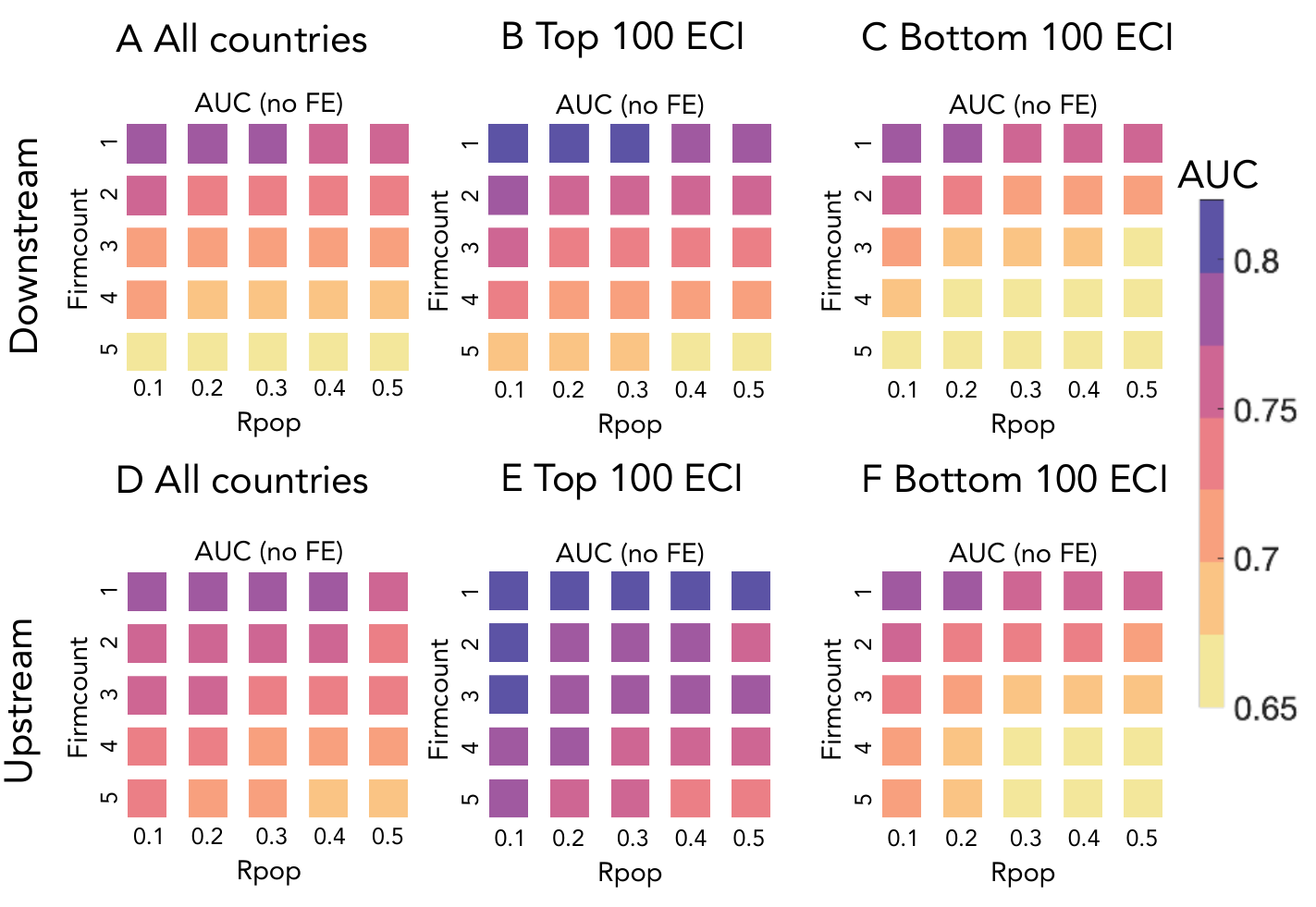}
  \caption{10 year growth regressions: all countries, high ECI countries, low ECI countries.}
  \label{SIfig1a}
\end{figure}

\newpage
\begin{figure}[t!]
  \centering
  \includegraphics[width=1\textwidth]{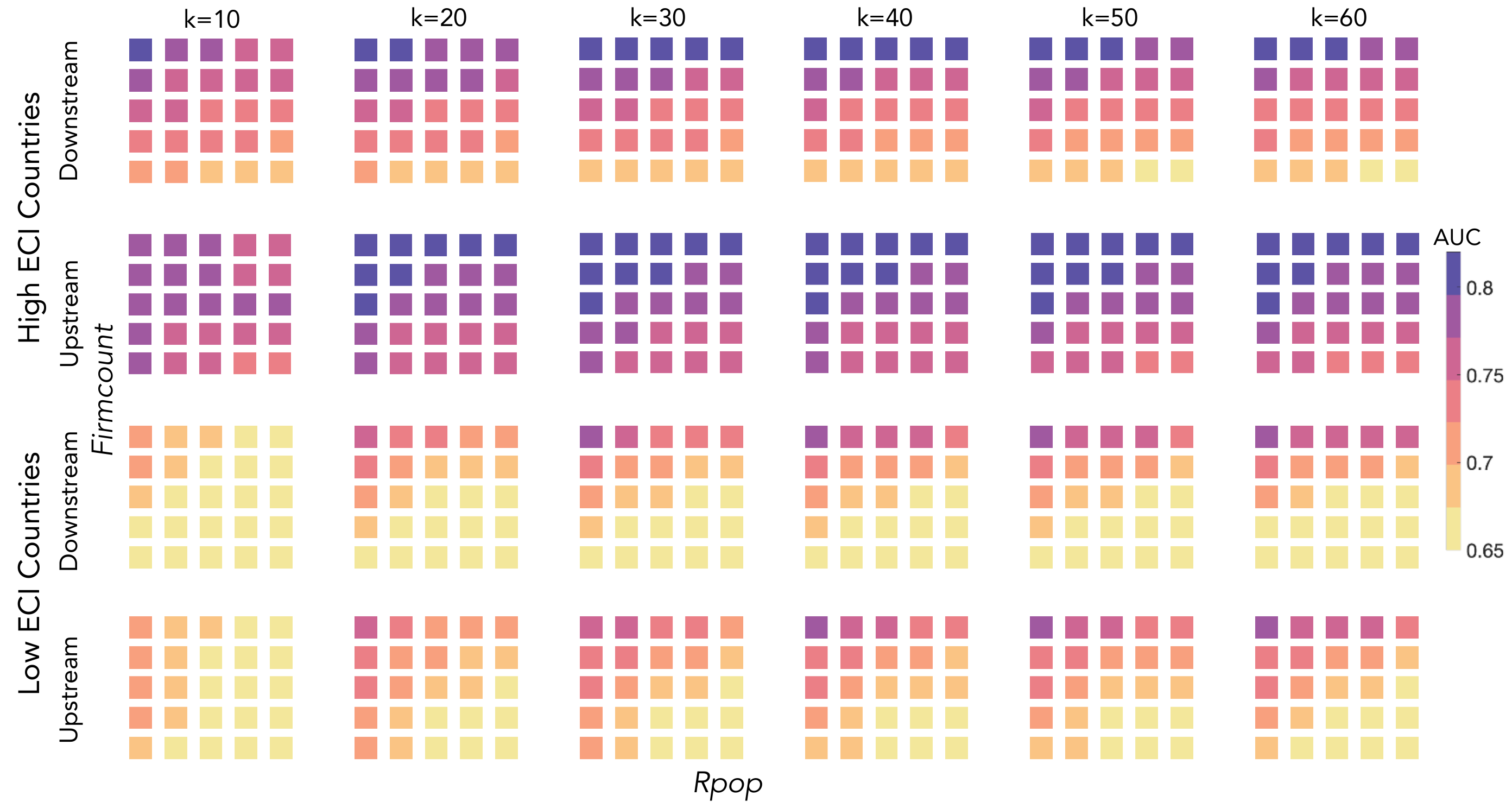}
  \caption{Variation in density parameter k for high and low ECI country sets. We see the AUC peaks around k=40 for all cases (more darker squares).}
  \label{SIfig1}
\end{figure}

\begin{figure}[t!]
  \centering
  \includegraphics[width=1\textwidth]{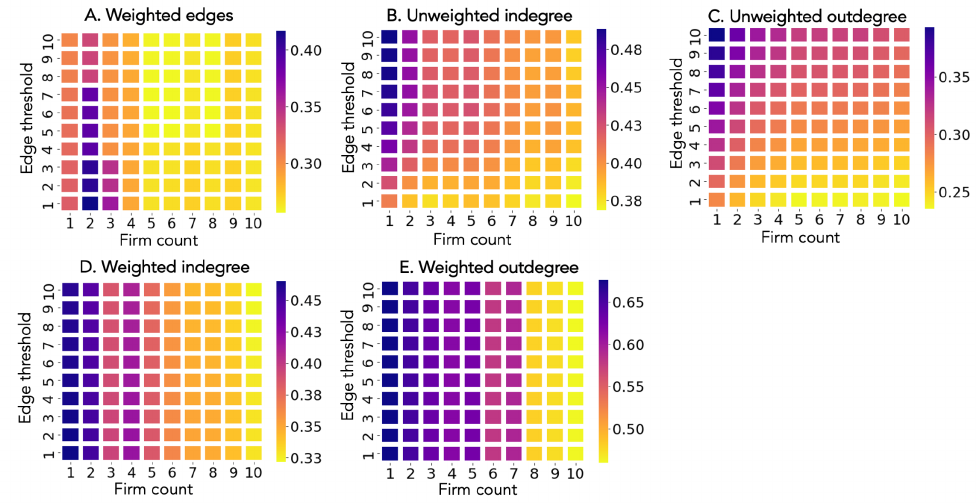}
  \caption{Our production network achieves a correlation in the range of 0.23 - 0.7 when compared to IO data (Eurostat 2022). We vary the edge and firm count thresholds of our network between 1-10 and correlate each instance of the network to the IO flows. Results are presented for correlations of the [A] weighted edges, [B-C] unweighted degrees (in/outdegrees), [D-E] weighted degrees (in/outdegrees) with the IO matrix.
}
  \label{SIfigIO}
\end{figure}

\end{document}